\documentclass[preprint,12pt]{elsarticle}

\usepackage[T1]{fontenc}
\usepackage[utf8]{inputenc}
\usepackage{amsmath,amssymb}
\usepackage{siunitx}
\usepackage{booktabs}
\usepackage{array}
\usepackage{longtable}
\usepackage{tabularx}
\usepackage{xltabular}
\newcolumntype{L}{>{\raggedright\arraybackslash}X}
\usepackage{graphicx}
\usepackage{textcomp}
\usepackage{lineno}
\usepackage{listings}
\lstdefinestyle{yamlspec}{basicstyle=\ttfamily\footnotesize,
  columns=fullflexible, keepspaces=true, frame=lines, framesep=5pt,
  aboveskip=1.1\baselineskip, belowskip=1.1\baselineskip, xleftmargin=1em}

\graphicspath{{./}}

\journal{Energy Conversion and Management}

\begin{document}

\begin{frontmatter}

\title{Automated generation of experimentally validated digital twins for
desiccant-based low-dew-point air-conditioning systems from declarative
topology specifications}

%% 저자 순서: Joo, Han, Oh, Bang, Kim (2026-07-09 확정)
%% 공동 교신저자: Joo, Kim
\author[kier,ust]{Younghwan Joo\corref{cor1}}
\ead{yhjoo@kier.re.kr}
\author[kier]{Jeonghoon Han}
\author[kier]{Sang Hyun Oh}
\author[kier]{Soosik Bang}
\author[kier,ust]{Sung-il Kim\corref{cor2}}
\ead{praygod@kier.re.kr}

\affiliation[kier]{organization={Energy Efficiency Research Division, Korea
  Institute of Energy Research},
  addressline={152 Gajeong-ro, Yuseong-gu},
  city={Daejeon},
  postcode={34129},
  country={Republic of Korea}}

\affiliation[ust]{organization={Energy Engineering, University of Science \&
  Technology},
  addressline={217 Gajeong-ro, Yuseong-gu},
  city={Daejeon},
  postcode={34129},
  country={Republic of Korea}}

\cortext[cor1]{Corresponding author}
\cortext[cor2]{Corresponding author}

\begin{abstract}
In battery manufacturing, the low-dew-point air conditioning of dry rooms is
among the largest energy consumers, and a physics-based digital twin offers
insight for operating-point optimization beyond the installed monitoring
points. Building one and calibrating it to field data each demand distinct
expertise, which limits industrial uptake. We present a framework that
generates a dynamic digital twin of an HVAC system from a declarative topology
specification, concise enough to draft from a natural-language plant
description, compiled against a purpose-built physical component library with
wiring, solver, and telemetry synthesized automatically. The models carry
equipment-level physics: the desiccant wheel couples heat and mass transfer
through an interchangeable sorption-isotherm component, so an undisclosed
commercial sorbent is calibrated as an effective isotherm rather than asserted
as a material. For experimental validation we built an industrial-grade,
ten-component low-dew-point system whose commercial desiccant-wheel unit holds a chamber near
$-40\,^{\circ}$C frost point, and operated it in both dehumidification and
bypass regimes. Generation reached a runnable model fifteen times faster than
expert manual construction, and a single parameter set, fitted only to three
closed-loop humidity nodes, predicts the bypass regime within
$0.1\,^{\circ}$C, the reactivation-heater power within 5\%, and measured
input-step responses. Identifiability analysis shows why this is prediction,
not fitting: ordinary operating points constrain only one parameter
combination, and the deep-dry equilibrium level of the recirculating loop
supplies the missing signal. The framework shortens the path from plant
description to measurement-validated twin; its criteria-based calibration is a
step toward twins calibrated, not only constructed, automatically.
\end{abstract}


\end{frontmatter}

\section{Introduction}\label{sec:introduction}

In the cleanrooms and drylabs of battery manufacturing, where product quality hinges on tight humidity control, low-dew-point air conditioning is among the largest energy consumers, on the order of 43\% of the energy of battery cell manufacturing \cite{yuan.etal2017ManufacturingEnergy,guan.etal2021OnsitePerformance,ma.etal2024OnsiteMeasurement}. Finding better operating points for these systems would save a large part of that energy, but the search must evaluate operating conditions the plant has never run, where no model, physical or data-driven, can promise accuracy in advance. A physics-based digital twin, calibrated against a modest set of field measurements, is the reasonable instrument for that search: it proposes operating candidates constrained by conservation laws and equipment physics rather than by the reach of the operating records, and each candidate the plant then tries returns measurements that tighten the calibration \cite{rajulapati.etal2022IntegrationMachine,wang.etal2023PhyllisPhysicsInformed,zhang.zhao2023DigitalTwin}.

Building such a twin is the hard part. It takes modeling expertise that is scarce outside specialist groups, and a model built carefully for one unit describes only that unit. HVAC is not a single system but dozens to hundreds of variants across a site, shaped by purpose from general comfort conditioning to industrial dehumidification, so hand-building a twin for each installation does not scale. Existing work on HVAC digital twins reflects this: fault detection, optimization, and online updating all run on top of an \emph{already-built} model, and the construction of the model itself stays a black box \cite{chen.etal2022DigitalTwins,hosamo.etal2022DigitalTwin,hosamo.etal2022DigitalTwina,abrazeh.etal2023VirtualHardwareintheLoop}. The bottleneck is the model, not its use.

This paper asks whether the twin can instead be generated from a declarative description of the system. If a plant's topology is stated declaratively, a physical model can be synthesized from it automatically; and the specification itself can be authored from a field engineer's natural-language account of the system together with a handful of P\&IDs, with a large language model (LLM) bridging the two. The pathway does not depend on who drives it, an engineer or an automated agent. What this paper implements and validates is the path from specification to twin: a physical twin of a complete HVAC system, resolved to equipment-level physics, generated from a declarative topology specification, that holds up against measurements from the system it represents. The natural-language front end enters once, as an end-to-end demonstration in the case study.

Automated model generation is not itself new. It has matured along two branches: one generating building energy models from BIM or building-stock data \cite{remmen.etal2017TEASEROpen,andriamamonjy.etal2018AutomatedIFCbased,sturzenegger.etal2014BRCMMatlab,jansen.etal2025OpensourceFramework,wetter.etal2021ModelicajsonTransforming}, the other generating digital twins of industrial process plants from P\&ID and asset data \cite{martinez.etal2018AutomaticGeneration,sierla.etal2022RoadmapSemiautomatic}. Neither reaches the physics an HVAC twin needs for optimization, which sits a level below the flowsheet of unit operations: a desiccant wheel is at once a heat exchanger, a moisture exchanger, and an adsorption process, coupling a nonlinear sorption isotherm, finite sorption kinetics, and carryover between its rotating zones. This is not a matter of difficulty but of what the application requires to be useful. Desiccant modeling at this fidelity is well established \cite{ge.etal2008ReviewMathematical,zhang.etal2003SimulationStudy,woods.kozubal2013DesiccantenhancedEvaporative,guan.etal2019ExperimentalNumerical}, but each study characterizes a single, known sorbent by hand, whereas the sorbent in a commercial rotor is typically undisclosed. No prior work generates physical models of complete HVAC systems with equipment-level physics from a declarative specification, at this depth and across commonly used equipment, and validates the result against an operating industrial system whose sorbent is unknown.

We develop the framework on an operating low-dew-point desiccant air-handling system at KIER, built around a Munters ML420 desiccant wheel of undisclosed sorbent. Within the framework the sorbent is an interchangeable component; for this system we calibrate an effective isotherm against measurement, keeping the calibration minimal and accounting for the one genuine unknown rather than tuning the model point by point. A single parameter set then reproduces the measured steady-state behavior across the wheel's active and bypassed regimes, so the validation tests the generated physics rather than a curve fitted to isolated data, a distinction we make precise in Section~\ref{sec:discussion}.

The paper's two contributions carry equal weight: a generated model earns the name of twin only when it answers to measurement, so construction and validation enter as stages of one pipeline. The first is the generation framework, which from a declarative YAML specification synthesizes a Simulink model, supplying a standard psychrometric fluid-property bus, a reusable component block library designed to span commonly used HVAC equipment, automatic wiring with mass- and energy-balance closure, solver configuration, and telemetry-tap insertion, and we quantify the construction effort it saves against manual model building. We do not automate the authoring of the underlying component physics; what the framework makes routine and reproducible is the composition of validated equipment-level physics from a declarative description. The second contribution is the experimental validation just described, on an industrial system with unknown sorbent, where a single parameter set reproduces the measured behavior across regimes and an identifiability analysis reports which parameters the data actually constrain. The declarative specification that drives the framework is also the form in which an autonomous agent could construct a twin on demand, a direction we pursue in future work.

\section{Automated model-generation framework}\label{sec:framework}

\begin{figure*}[htbp]
\centering
\includegraphics[width=\textwidth]{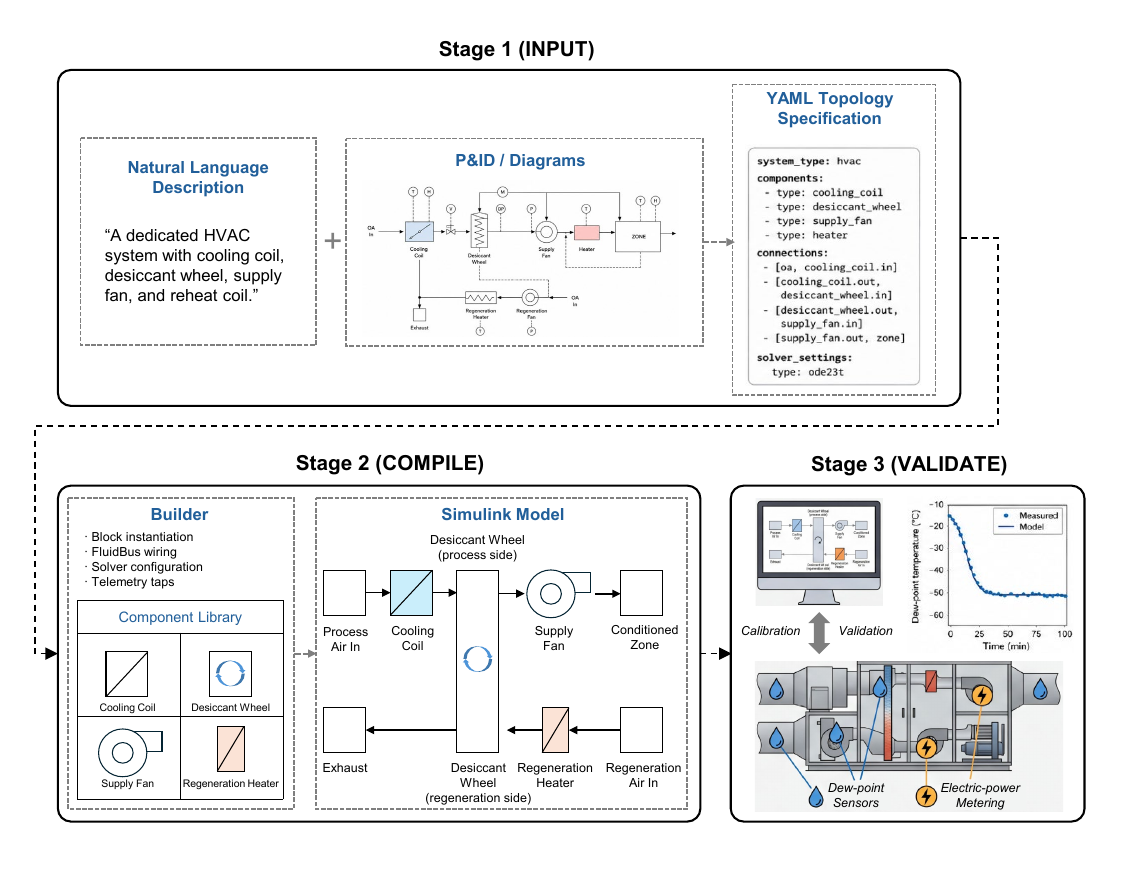}
\caption{Overview of the automated generation and validation framework. A declarative YAML topology specification is compiled against a reusable component library sharing a standardized moist-air bus interface (FluidBus). The builder synthesizes a runnable Simulink model (block instantiation, bus wiring, solver configuration, and telemetry taps), which is then calibrated and validated against an operating industrial low-dew-point desiccant air-handling system using dew-point and electric-power measurements.}
\label{fig:overview-pipeline}
\end{figure*}

Figure~\ref{fig:overview-pipeline} shows the framework end to end. A topology specification written in YAML declares the system: the components it contains, how their air streams connect, the operating inputs that drive them, and the signals to be observed. A builder compiles this specification against a library of physical component blocks into a runnable Simulink model, configures its solver, and inserts telemetry taps, and the generated model then enters the calibration and validation workflow of Sections 4 and 5. The pipeline also includes a specification-authoring stage in front of the builder: the YAML can be written directly by an engineer, or produced from a natural-language description of the system and a few P\&IDs by an LLM, supported by surveyed sizing ranges for common industrial HVAC equipment when exact values are unavailable. We demonstrate this front stage once, end to end, in the case study; the quantitative claims of the paper concern the path from specification to validated model.

The framework rests on a single idea: it reduces the construction of a system digital twin with equipment-level physics from a modeling problem to a composition problem. Every component exchanges the same moist-air state over a standardized bus, so a system is fully specified by naming its blocks and their connections, and everything else (wiring, balance closure, solver setup, instrumentation) follows mechanically from the specification. The point of this reduction is to guide whoever drives the generation, a person authoring a specification or the code assembling the model from it, so that neither loses its way nor has to solve a harder problem than the application requires. The remainder of this section presents the framework in that order: the interface that performs the reduction (2.1), the component vocabulary built on it (2.2), the synthesis of a model from a specification (2.3), the residual difficulties that the framework absorbs rather than passes on (2.4), and the range of system structures the same machinery covers (2.5).

\subsection{Standard moist-air bus interface}

All air-stream connections in the framework carry a single bus signal, the FluidBus, holding four state variables in SI units: temperature (K), pressure (Pa), mass flow rate (kg/s), and relative humidity (0--1). Any derived psychrometric quantity (humidity ratio, dew point, enthalpy) is computed inside the blocks from this state through a shared set of property functions following standard psychrometric formulations \cite{herrmann.etal2009ThermodynamicProperties}, so that all components evaluate moist-air properties consistently and closed air loops conserve mass and energy across block boundaries.

The restriction to four variables is a design decision rather than a simplification of convenience. Because every outlet and every inlet speak the same contract, connecting any two ports is well defined without further judgment: the builder does not need to reason about which quantities a particular pairing exchanges, and the author of a specification does not need to know how any block represents its internal physics. Composition becomes mechanical, which is precisely what makes automatic wiring (by code or, prospectively, by an agent) a tractable problem. Component models of heterogeneous origin (Simscape blocks, custom physical models) are wrapped behind this interface, with non-stream quantities such as heat and moisture loads exposed as separate scalar ports.

\subsection{Component block library}

The component vocabulary is catalogued in a machine-readable library declaration that records, for every block type, its port signature on the bus and its mask-parameter schema (Table~\ref{tab:framework-library}). The library spans the equipment that commonly composes industrial air-handling systems: desiccant wheels, cooling coils, heaters, humidifiers, mixers and flow splitters, and conditioned spaces, a vocabulary in the tradition of component libraries for building simulation \cite{wetter.etal2013ModelicaBuildings}, centered here on industrial dehumidification. Air flow rates are imposed through the specification (source flow rates at the boundaries and split ratios at the dividers) rather than resolved by the components. The governing physics of the blocks central to this study are presented in Section 4; standard components follow established formulations.

The mask parameters of a block are the information a specification must supply to instantiate it in a new system, and the library deliberately keeps this set small. Each additional parameter is a value the specification author, again a person or an agent, must know or guess, and any parameter that cannot be traced to a datasheet or a measurement ultimately becomes a calibration burden. The blocks therefore expose the minimal parameter set that preserves the depth needed for operating-point and energy prediction, a restraint that Section 4 revisits when the sources of every model input are accounted for.

The sorbent of the desiccant wheel illustrates the same principle applied to material properties. Rather than hard-coding one material, the wheel block treats the sorption isotherm as an interchangeable component: isotherm models for a set of candidate sorbents are registered in a common lookup interface and selected per instance through a mask parameter. How this supports systems whose sorbent is undisclosed (the situation of most commercial rotors, including the one in our case study) is developed in Section 4.

\subsection{From declarative specification to executable model}

A topology specification declares, in order, the model metadata, the boundary air sources and their conditions, the component instances with their mask overrides and control inputs, the flow and control connections, the observed outputs, any special handling such as feedback points, and the solver settings. The specification is the single source of truth for the model: every block, wire, and parameter value in the generated Simulink diagram derives from it, and regenerating from the same file reproduces the model identically. The excerpt below, taken from the case-study specification of Section~3, shows this anatomy; the complete diagram it generates, and its manually constructed counterpart, are compared in Section~5.1.

\begin{lstlisting}[style=yamlspec]
# topo14 specification (excerpt)
metadata:
  model_name: Topo14_KIER_NIR_OP3_DeepDry

components:
  - name: Rotor
    block_type: Rotor_Bus
    mask_parameters:
      desiccant_material: Zeolite4A_E18
      D_rotor: 0.65            # m
      k_LDF: 0.04              # m^3 kg^-1 s^-1
      ab_frac: 0.667
  - name: AHU_Cooler
    block_type: Cooler_Bus
  # ... 7 more components

connections:
  flow_connections:
    - { from: Mixer/1, to: Rotor/1, signal_type: FluidBus }
    - { from: Rotor/1, to: Splitter/1, signal_type: FluidBus }
    # ...

solver:
  type: ode23t
  stop_time: 12000.0
\end{lstlisting}

Synthesis begins with validation. The builder checks the specification structurally against the library catalogue (block types, port indices, telemetry targets) and physically against admissible ranges: temperatures must fall within a plausible HVAC envelope in kelvin, pressures near atmospheric in pascals, humidities within the unit interval, and the selected sorbent must be compatible with the declared regeneration temperature. Consistency of component sizing with the imposed flows is a precondition for a convergent simulation, so ill-posed specifications are rejected before any model is built, with diagnostics that identify the likely authoring error, most commonly a unit mistake, for which the message proposes the SI conversion. Only a specification that passes these checks proceeds to synthesis: components are instantiated from the library, bus connections are wired, unused ports are terminated, feedback points receive the treatment of Section~\ref{subsec:closed-loop}, the solver is configured, telemetry taps are inserted, and a build report is emitted for downstream analysis.

\subsection{Closed-loop topologies and solver handling}\label{subsec:closed-loop}

Recirculating air paths, the rule rather than the exception in industrial dehumidification, close algebraic loops through the model. The specification marks each feedback point, and the builder breaks the loop with an initialized memory state carrying a declared starting condition for the bus variables, so that closed-loop models initialize and integrate reliably. Solver selection is likewise part of the specification and its defaults: stiff integrators suited to the topology class are configured automatically, so that neither the specification author nor the generated model's user tunes solvers by hand. These are exactly the difficulties that historically made closed-loop HVAC models expert work, and the framework's role is to absorb them rather than pass them on.

Telemetry taps declared in the specification are inserted as a communication subsystem at build time, making every generated model twin-ready; online operation against the physical system is beyond this paper's scope and is pursued separately.

\subsection{Generality}

The same machinery covers a range of system structures, which we illustrate with four representative topologies rather than an exhaustive gallery: a once-through outdoor-air handling unit (open topology), a recirculating closed loop through a desiccant wheel (the feedback structure of Section~\ref{subsec:closed-loop}), an isolated rotor model used for single-component analysis, and a bypass variant in which the wheel and its regeneration path are removed from the air path by a local edit to the specification. In total, fourteen topology families have been generated with the framework during this work. The bypass case deserves note: the two operating configurations validated in Section 5 (dehumidification and bypass) are not two hand-built models but two specifications differing in a few lines, which is itself evidence of what declarative generation buys. The framework is designed for this structural generality; its experimental validation in this paper is on one industrial system, in two operating regimes, as Sections 3--5 develop.

%% [Table:framework-library] — authored from topologies/library_blocks.yaml
%% 모든 블록은 표준 moist-air 버스(FluidBus: T, P, m_dot, RH)를 공유. 아래는 버스 외 보조 포트와 주요 마스크 파라미터.
\begin{table}[htbp]
\centering
\caption{Component block library: physical blocks, their port signatures on
the standard moist-air bus (FluidBus: $T$, $P$, $\dot m$, RH), and the principal
mask parameters a specification must supply.}
\label{tab:framework-library}
\begin{tabularx}{\textwidth}{@{}l >{\raggedright\arraybackslash}p{0.30\textwidth} L@{}}
\toprule
Block & Bus ports (auxiliary ports) & Principal mask parameters \\
\midrule
Desiccant wheel & process in/out, regeneration in/out (rotation speed) &
sorbent material, $D_r$, $L_r$, $k_{LDF}$, adsorption-sector fraction $f_{ab}$
(three-sector purge variant available) \\
\addlinespace
Cooling coil & air in/out, chilled-water side (heat load, condensate) &
tube count, rows, fin area, coolant supply temperature \\
\addlinespace
Heater & air in/out (heat duty / setpoint) &
$UA$, fin efficiency \\
\addlinespace
Humidifier & air in/out (water addition) &
supply-water temperature (target humidity via input) \\
\addlinespace
Mixer & two air inlets, one outlet &
--- (adiabatic mixing) \\
\addlinespace
Flow splitter & one inlet, two outlets &
split ratio \\
\addlinespace
Conditioned space & supply in, return out (heat load, moisture source) &
volume $V_{room}$, leakage/return conductances $K_{leak}$, $K_{return}$ \\
\bottomrule
\end{tabularx}
\end{table}

\section{Case-study system}\label{sec:case-study}

\begin{figure}[htbp]
\centering
\includegraphics[width=\textwidth]{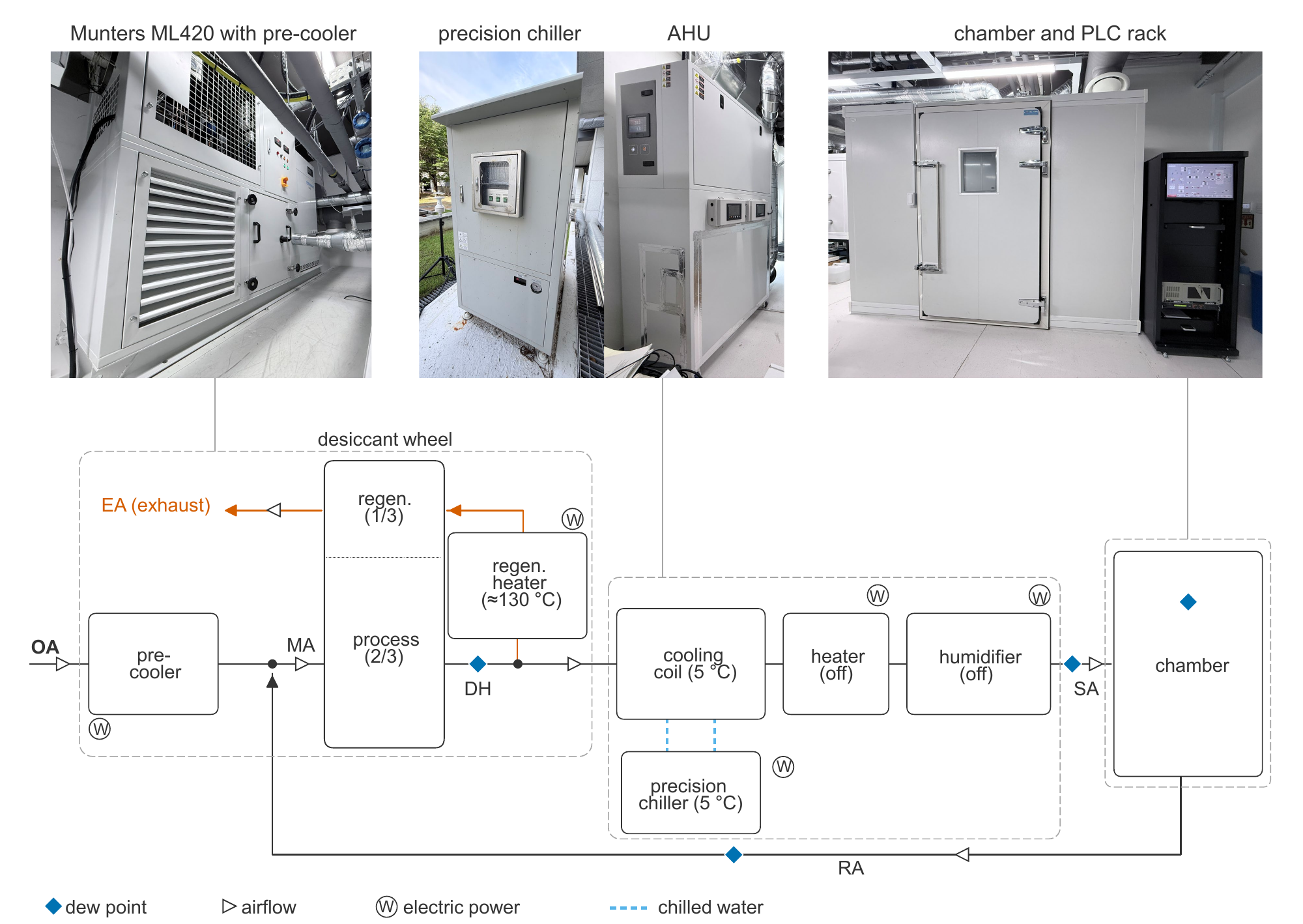}
\caption{The industrial low-dew-point air-handling system used as the validation testbed: schematic of the air paths, the two cold sources, and the instrumentation, with photographs of the principal equipment linked to the corresponding components. Details are given in Section 3.}
\label{fig:case-system}
\end{figure}

We develop and validate the framework on an operating low-dew-point air-handling system at the Korea Institute of Energy Research (KIER), which conditions a process chamber, the type of conditioned space that, in production settings, would house humidity-sensitive manufacturing steps (Fig.~\ref{fig:case-system}). The system is industrial in kind rather than a laboratory mock-up: its desiccant unit, air-handling unit, and chamber are commercial equipment operated by the facility's PLC, and it was running its normal duty throughout the measurement period.

\subsection{System description}

Dehumidification is provided by a Munters ML420 desiccant wheel unit with two sectors (process and reactivation, with no purge sector), regenerated at approximately $130\,^{\circ}$C and rated at 420 m$^3$/h of process air against 155 m$^3$/h of reactivation air. The wheel sits inside a recirculating loop: outdoor make-up air, pre-cooled by a unit packaged with the desiccant system, mixes with return air from the chamber, and the mixed stream passes the process sector of the wheel; the dried stream splits between supply air to the chamber and a branch that is heated and fed back through the reactivation sector in counter-flow. The cooling coil of the air-handling unit downstream is served by chilled water from a dedicated precision chiller, while the pre-cooler is a self-contained packaged refrigeration unit, holding its outlet near $8\,^{\circ}$C under local control (Section~\ref{sec:instrumentation}). The reference airflow structure, established by mass-balance closure of the measured airflows (Section 4), maintains mixed, exhaust, and supply flows in the ratio MA : EA : SA = 3 : 1 : 2 (421, 171, and 250 m$^3$/h respectively, with 232 m$^3$/h of outdoor air and 189 m$^3$/h of return air) (Table~\ref{tab:case-specs}).

\begin{figure}[htbp]
\centering
\includegraphics[width=\textwidth]{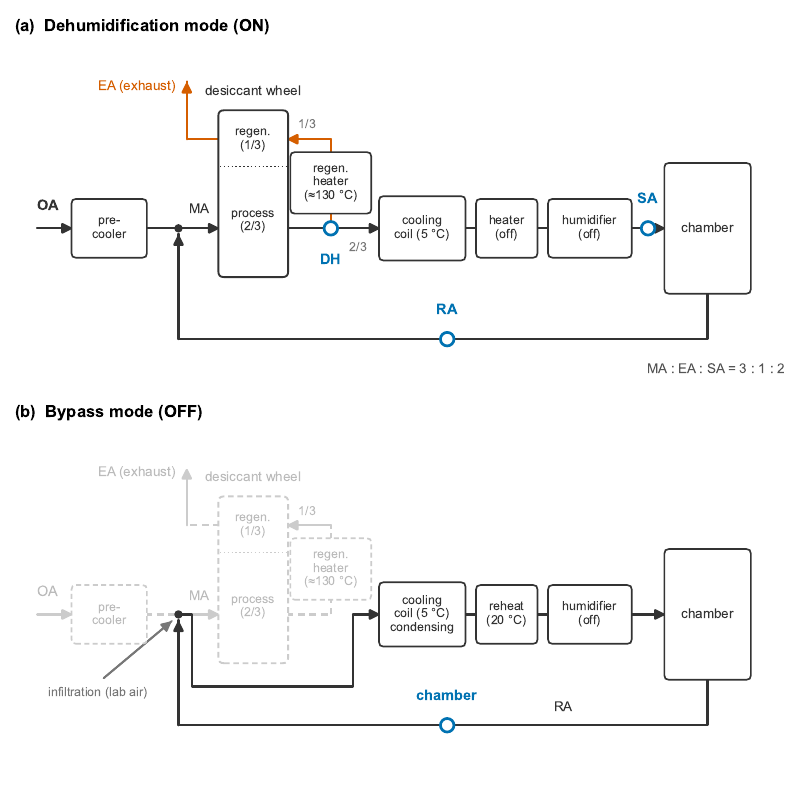}
\caption{The two operating configurations used for validation. (a) Dehumidification mode (ON): recirculating closed loop through the desiccant wheel, with a one-third regeneration split downstream of the process outlet fed back through the regeneration sector in counter-flow. (b) Bypass mode (OFF): the wheel, pre-cooler, and regeneration path are inactive (greyed); the recirculating air routes around the wheel and the chamber is conditioned by the condensing cooling coil alone, with laboratory-air infiltration as the moisture load. Circles mark the measurement nodes used in Section 5.}
\label{fig:case-two-op}
\end{figure}

The facility operates the system in two regimes as a matter of routine. In dehumidification mode (ON), the full closed loop through the wheel is active and the chamber reaches deep-dry conditions, with dew points near $-40\,^{\circ}$C at the wheel outlet. In bypass mode (OFF), the wheel, pre-cooler, and reactivation path are inactive; the recirculating air routes around the wheel, and the chamber is held by the condensing cooling coil alone against laboratory-air infiltration, at dew points near $+5\,^{\circ}$C (Fig.~\ref{fig:case-two-op}). These two regimes form the validation pair used throughout Sections 4 and 5; they also span the two standard dehumidification routes, desiccant and condensing \cite{ge.wang2020ExergyAnalysis}. They are attractive as a pair precisely because they are operationally natural rather than contrived for the study. On the modeling side, the two configurations are not two hand-built models but two topology specifications differing in a few lines, as noted in Section 2.5.

\subsection{Instrumentation}\label{sec:instrumentation}

All measurements come from the facility's PLC, accessed read-only through its monitoring interface, which provides live values and logged histories at intervals of a few seconds. Four groups of channels are used in this study. Polymer/QCM dew-point transmitters (SUTO S220; $-100$ to $+20\,^{\circ}$C Td measurement range, $\pm 2\,^{\circ}$C Td accuracy over $-70$ to $0\,^{\circ}$C Td) are installed at the supply air, the wheel process outlet, and the return air; a fourth dew-point channel inside the chamber showed a consistent high bias against the surrounding sensors during commissioning checks and is therefore excluded from the validation set, the chamber condition being represented by the return-air measurement instead. Dry-bulb temperatures are available at the principal points from the equipment controllers, including the air-handling unit supply temperature and the chilled-water supply temperature (Pt100). Airflow stations report the volumetric flows of the principal streams. Electric power is metered per device (among others the reactivation heater, the supply-air compressor and fans, and the chiller), alongside aggregate channels. The pre-cooler exposes no monitoring points beyond its electric-power channel; its outlet temperature is regulated to a fixed setpoint at the equipment panel and is not logged.

The validation data are taken from steady operating windows in each regime, one in dehumidification mode and one in bypass mode, identified from the logged histories; the window statistics and their use in the validation are given in Section 5. How the raw channels are conditioned into the quantities the validation uses, including the mass-balance closure of the airflow readings and the interpretation of the power channels, is part of the validation methodology and is treated in Section 4.

\subsection{Characteristics that shape the validation methodology}

Three characteristics of this system determine how the generated model can be validated, and Section 4 builds the methodology around them.

First, the sorbent of the ML420 wheel is not disclosed by the manufacturer, which is the usual situation for commercial desiccant rotors. The wheel model therefore cannot be parameterized from a published isotherm of a known material; instead, the framework's interchangeable-isotherm mechanism (Section 2.2) is used to calibrate an effective isotherm from measurement, a procedure developed in Section 4 together with the analysis of which parameters the data actually constrain.

Second, the operating range spans dew points from $+5\,^{\circ}$C down to $-40\,^{\circ}$C. In the deep-dry range, relative-humidity sensing loses resolution (the quantities of interest live in the last fraction of a percent of relative humidity), so agreement is assessed on dew-point temperature directly, and dew point is adopted as the primary validation metric throughout (Section 4).

Third, the system runs as a recirculating loop, so any error in a single component's outlet state is fed back to its inlet and accumulated pass over pass rather than washed out. Validation therefore cannot stop at single-pass component checks: the generated model must be run to closed-loop equilibrium and judged there, which motivates the drift-based equilibrium criteria of Section 4 and, in Section 5, turns the loop's sensitivity into an asset: the closed-loop response separates parameter variants that single-pass observations cannot distinguish.

%% [Table:case-specs] — authored from data (topo14 yaml + ML420 datasheet + PLC tags)
%% 수치 검증 완료 (2026-07-10, 원천 대조): ML420 420/155 m³/h = 제품시트 p2 판독 일치.
%% ⚠ O2 반영 (2026-08-05, A안): "Total electrical rating 4.57 kW" 행 삭제 — 카탈로그 표준구성 정격이
%%   사이트 130 °C 2뱅크 히터 실측(5.171 kW)과 표면 모순을 유발(Oh 지적). 논문 주장 비의존 수치라 제거.
%%   130°C = ref yaml automatic + T5 문서. 풍량 421/171/250/232/189/61 = OP3 질량폐합(§4 표와 동일,
%%   topo14 yaml OA 232·RA 189·bleed 61 정합). 5°C 항온수조·8°C 프리쿨러 = ref yaml setpoints. ~15s = 전력창 48샘플/12분.
%%   SUTO S220 = 업체 견적서로 모델명 확인 (저자 확인 2026-07-10); 사양(-100~+20 Td, ±2 over -70~0)은 SUTO 공개 데이터시트.
%% → Table 2 전 수치 검증 종결 (2026-07-10).
\begin{table}[htbp]
\centering
\caption{Case-study system and instrumentation summary: desiccant unit ratings
(Munters ML420), reference airflow structure, cooling provisions (a
chilled-water air-handling coil and a packaged pre-cooler), and measurement
channels (dew point, airflow, per-device electric power).}
\label{tab:case-specs}
\begin{tabularx}{\textwidth}{@{}l L@{}}
\toprule
Item & Value \\
\midrule
\multicolumn{2}{@{}l}{\textit{Desiccant unit (Munters ML420)}}\\
\quad Sectors & Process / reactivation (no purge) \\
\quad Process air flow & 420\,m$^3$\,h$^{-1}$ (rated) \\
\quad Reactivation air flow & 155\,m$^3$\,h$^{-1}$ (rated) \\
\quad Regeneration temperature & $\approx$130\,$^{\circ}$C \\
\midrule
\multicolumn{2}{@{}l}{\textit{Reference airflow structure (mass-balance closure)}}\\
\quad Ratio MA : EA : SA & 3 : 1 : 2 \\
\quad Mixed / exhaust / supply (MA/EA/SA) & 421 / 171 / 250\,m$^3$\,h$^{-1}$ \\
\quad Outdoor / return (OA/RA) & 232 / 189\,m$^3$\,h$^{-1}$ \\
\quad Chamber bleed & 61\,m$^3$\,h$^{-1}$ \\
\midrule
\multicolumn{2}{@{}l}{\textit{Cooling provisions}}\\
\quad AHU cooling coil & Chilled water from a dedicated precision chiller ($\approx$5\,$^{\circ}$C) \\
\quad Outdoor-air pre-cooler & Self-contained packaged unit, outlet $\approx$8\,$^{\circ}$C \\
\midrule
\multicolumn{2}{@{}l}{\textit{Instrumentation}}\\
\quad Dew point & SUTO S220 transmitters at SA, wheel outlet (DH), RA \\
 & ($-100$ to $+20\,^{\circ}$C\,T$_\mathrm{d}$; $\pm2\,^{\circ}$C over $-70$ to $0\,^{\circ}$C\,T$_\mathrm{d}$) \\
\quad Temperature & Controller channels incl.\ AHU supply, chilled-water supply (Pt100) \\
\quad Airflow & Volumetric-flow stations on the principal streams \\
\quad Electric power & Per-device metering (kW, $\approx$15\,s logging) \\
\bottomrule
\end{tabularx}
\end{table}

\section{Physical component models and validation methodology}\label{sec:validation}

The models generated by the framework combine two custom component models built for this work (the desiccant wheel and the conditioned chamber) with standard moist-air components. This section presents the governing physics of the custom blocks (4.1, 4.2), summarizes the standard components (4.3), and then sets out the methodology by which the generated system model is calibrated and judged against measurement (4.4).

\subsection{Desiccant wheel model}

\begin{figure}[htbp]
\centering
\includegraphics[width=\textwidth]{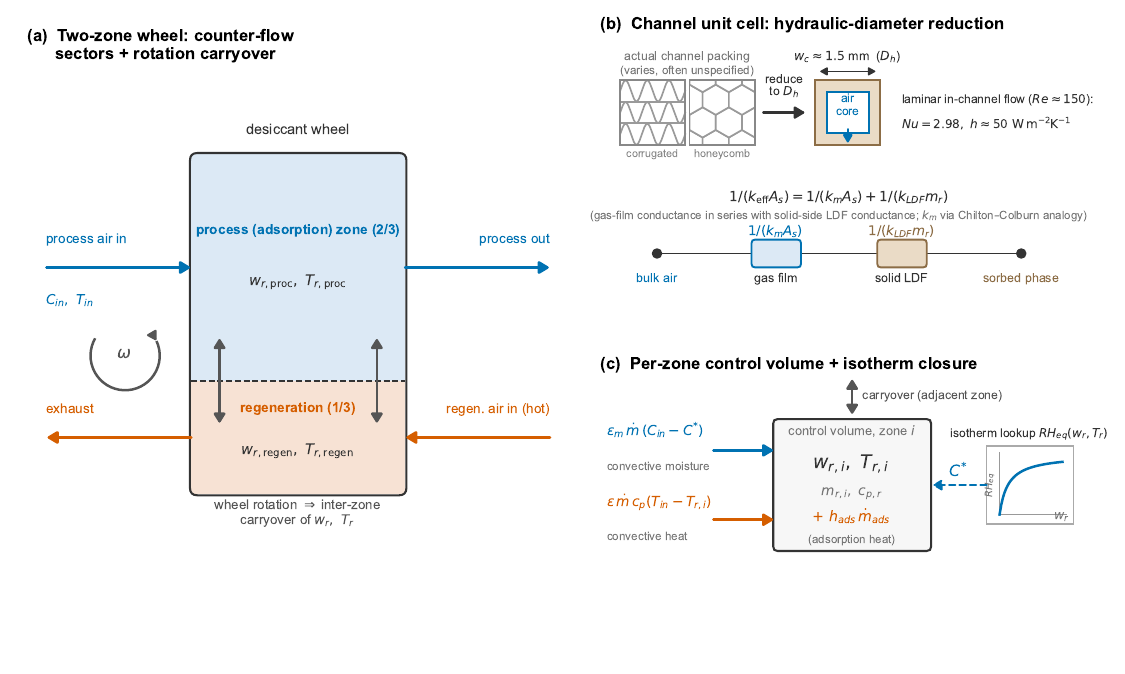}
\caption{Two-zone lumped model of the desiccant wheel. (a) Wheel geometry: process (2/3) and regeneration (1/3) sectors with counter-flow air streams; wheel rotation exchanges matrix mass and energy between zones (carryover). (b) Channel transfer model: the channel is reduced to an equivalent square channel of hydraulic diameter $D_h = w_c \approx 1.5$ mm, and the gas-film conductance from the Chilton--Colburn analogy combines in series with the solid-side linear-driving-force (LDF) conductance in the form of Eq.~\ref{eq:rotor-ldf}. (c) Per-zone state variables ($w_r$, $T_r$) and NTU-effectiveness heat/mass exchange closed by the sorption-isotherm lookup $RH_{eq}(w_r, T_r)$.}
\label{fig:rotor-model}
\end{figure}

The wheel is modeled as lumped zones corresponding to its sectors: two zones for the case-study unit, the process (adsorption) sector and the regeneration sector, with area fractions 2/3 and 1/3, and the same formulation extends to three-sector wheels that carry a purge zone, which the library supports as a variant block. Each zone carries two states, the moisture content of the desiccant matrix $w_{r,i}$ and its temperature $T_{r,i}$, where the zone index runs over the two sectors, $i \in \{\mathrm{proc}, \mathrm{regen}\}$, and $j$ denotes the other zone of the pair (Fig.~\ref{fig:rotor-model}). Each sector is represented by its mass-averaged state (the angular distribution within a sector is not resolved, in contrast to spatially resolved formulations \cite{sphaier.worek2004AnalysisHeat}), and wheel rotation enters explicitly as a carryover term: matrix mass and energy cross between zones at a rate proportional to the rotation speed $\omega$.

The moisture balance of zone $i$ reads

\begin{equation}
m_{r,i}\, \frac{dw_{r,i}}{dt} = \underbrace{\varepsilon_{m,i}\, \frac{\dot m_i}{\rho_{air}}\, (C_{in,i} - C^{*}_{i})}_{\dot m_{ads,i}} \;+\; \underbrace{\rho_{por} L_r \left(\tfrac{D_r}{2}\right)^{2} \tfrac{\omega}{2}\, (w_{r,j} - w_{r,i})}_{\text{carryover}}
\label{eq:rotor-moisture}
\end{equation}

where the air-side mass exchange is written in NTU-effectiveness form \cite{ruivo.angrisani2014EffectivenessMethod}, $\varepsilon_{m,i} = 1-\exp(-\mathrm{NTU}_{m,i})$ with $\mathrm{NTU}_{m,i} = k_{eff} A_{s,i}\, \rho_{air}/\dot m_i$, which keeps the exchange bounded by its thermodynamic limit at any flow rate and remains well-posed as $\dot m_i \to 0$. The driving potential is the difference between the inlet vapor concentration and the equilibrium concentration at the matrix surface,

\begin{equation}
C^{*}_{i} = \frac{p_{sat}(T_{r,i})\; RH_{eq}(w_{r,i},\, T_{r,i})}{R_w\, T_{r,i}}
\label{eq:rotor-eq-conc}
\end{equation}

where $RH_{eq}(w, T)$ is the sorption-isotherm closure, implemented as a two-dimensional lookup so that the sorbent is an interchangeable component of the model (Section 2.2); the isotherm family used for the case study is developed below.

The heat balance of zone $i$ carries the convective exchange, the rotational carryover, and the heat of adsorption:

\begin{equation}
m_{r,i}\, c_{p,r}\, \frac{dT_{r,i}}{dt} = \varepsilon_i\, \dot m_i c_{p,air}\, (T_{in,i} - T_{r,i}) \;+\; \rho_{por} L_r \left(\tfrac{D_r}{2}\right)^{2} \tfrac{\omega}{2}\, c_{p,r}\, (T_{r,j} - T_{r,i}) \;+\; h_{ads}\, \dot m_{ads,i}
\label{eq:rotor-heat}
\end{equation}

with $\varepsilon_i = 1-\exp(-\mathrm{NTU}_i)$, $\mathrm{NTU}_i = h A_{s,i}/(\dot m_i c_{p,air})$, and the sign convention that adsorption ($\dot m_{ads} > 0$) releases heat and desorption absorbs it. The specific heat of adsorption $h_{ads}$ is a sorbent property taken from the literature for each registered material.

The air-side outlet states follow from the same effectiveness factors,

\begin{equation}
C_{out,i} = C^{*}_{i} + (C_{in,i} - C^{*}_{i})\, e^{-\mathrm{NTU}_{m,i}}, \qquad
T_{out,i} = T_{r,i} + (T_{in,i} - T_{r,i})\, e^{-\mathrm{NTU}_{i}}
\label{eq:rotor-outlet}
\end{equation}

so that the heat of adsorption is assigned entirely to the matrix and reaches the air only through the raised matrix temperature $T_{r,i}$. In the low-flow limit the outlet states approach the matrix states, $C_{out,i} \to C^{*}_{i}$ and $T_{out,i} \to T_{r,i}$, which is what keeps the block well-posed when a stream is shut off.

\paragraph{Transfer coefficients} The channel microstructure of commercial wheels is often unspecified and varies between products (corrugated, honeycomb, and other matrix geometries), so the model reduces the channel to an equivalent square duct of hydraulic diameter $D_h = 4A/P = w_c$. At the case-study operating conditions the channel Reynolds number is of order $10^2$, well within the laminar regime; with the flow thermally fully developed over the wheel depth, the constant-wall-temperature Nusselt number for a square duct applies, $Nu = 2.98$ \cite{shah.london1978RectangularDucts}, giving the gas-film heat-transfer coefficient $h = Nu\, k_{air}/D_h \approx 50\ \mathrm{W\,m^{-2}K^{-1}}$. (The implemented correlation also covers developing and turbulent branches, but the operating point does not reach them.) The gas-film mass-transfer coefficient follows from the Chilton--Colburn analogy \cite{chilton.colburn1934MassTransfer},

\begin{equation}
k_m = \frac{h}{\rho_{air}\, c_{p,air}}\, Le^{-2/3}
\label{eq:rotor-chilton-colburn}
\end{equation}

and is combined in series with a solid-side linear-driving-force (LDF) coefficient $k_{LDF}$ \cite{glueckauf1955TheoryChromatography}, which lumps the intra-particle sorption kinetics. The two act on different bases, the gas film on the transfer area and the solid uptake on the sector's desiccant mass, so the combination is written between conductances: the gas-film conductance $k_m A_{s,i}$ in series with the solid-side conductance $k_{LDF}\, m_{r,i}$,

\begin{equation}
\frac{1}{k_{eff}\, A_{s,i}} = \frac{1}{k_m\, A_{s,i}} + \frac{1}{k_{LDF}\, m_{r,i}}
\label{eq:rotor-ldf}
\end{equation}

Because the driving potential here is a vapor concentration rather than a matrix loading, $k_{LDF}$ carries units of \si{\cubic\metre\per\kilogram\per\second}, and relates to the loading-based coefficient $k$ (\si{\per\second}) conventional in the LDF literature by $k_{LDF} = k\, \partial w/\partial C$; the two are numerically similar in magnitude for this sorbent but are not the same quantity. A gas-film-only formulation transfers moisture at whatever rate the film allows and, for this wheel, overpredicts single-pass uptake; the finite solid-side kinetics is what lets one model reproduce both the static depth and the closed-loop dynamics, as Section 5 quantifies. Geometry closes through the unit cell: face area $A_c = \tfrac{\pi}{4} D_{r}^2 f_{zone}$, channel velocity $V = \dot m/(\rho_{air} A_c \phi)$, and transfer area $A_s = 4\phi A_c L_r / w_c$ for porosity $\phi$.

\paragraph{Sorption isotherm and its calibrated family} Registered sorbents are described by their published isotherm models and compiled to the $RH_{eq}(w, T)$ lookup; for the zeolite family used here, the model is a triple-site Langmuir (TSL) with site affinities $b_k(T) = b_{0,k} \exp(E_k/RT)$ \cite{wang2020MeasurementsModeling}. Because the case-study sorbent is undisclosed, no library material can be asserted a priori; Section 4.4 describes the screening that selects a base material, and the calibration then acts on a one-parameter family of the base isotherm. The family rescales the site activation energies, $E_k' = s E_k$, while renormalizing the pre-exponentials,

\begin{equation}
b_{0,k}' = b_{0,k}\, \exp\!\left[\frac{E_k (1-s)}{R\, T_{ref}}\right]
\label{eq:rotor-escale}
\end{equation}

so that at $T = T_{ref}$ the scaled exponent $s E_k/R T_{ref}$ and the compensating $E_k(1-s)/R T_{ref}$ recombine to the original $E_k/R T_{ref}$: $b_k(T_{ref})$, and with it the isotherm at the process temperature, is pinned exactly, and the scale factor $s$ modifies only the isotherm's decline toward regeneration temperatures. The construction makes $s$ a pure desorption-side knob, structurally orthogonal to any calibration performed at process conditions (Fig.~\ref{fig:results-isotherm-escale}), a property that Section 4.4 relies on to keep the calibration ledger disjoint.

\subsection{Conditioned-chamber model}

Low-dew-point chambers are operated at positive gauge pressure so that leakage flows outward, which makes infiltration-versus-leakage physics the first-order moisture behavior of the room. The chamber model tracks dry air and water vapor separately,

\begin{equation}
\frac{dm_{dry}}{dt} = \dot m_{dry,in} - \dot m_{dry,out} - \dot m_{dry,leak}
\label{eq:room-dry}
\end{equation}

\begin{equation}
\frac{dm_{wv}}{dt} = \dot m_{dry,in} Y_{in} - \dot m_{dry,out} Y_{out} + \dot m_{source} - \dot m_{wv,leak}
\label{eq:room-vapor}
\end{equation}

with $Y$ the humidity ratio and $\dot m_{source}$ an internal moisture source that accepts scheduled scenarios. The chamber is treated as well-mixed, so the outgoing streams carry the chamber's own composition, $Y_{out} = m_{wv}/m_{dry}$ and $\dot m_{wv,leak} = \dot m_{leak} Y_{out}$. Dry-basis and total flows are related by $\dot m = \dot m_{dry}(1+Y)$ and the inventories by $m = m_{dry} + m_{wv}$, with composition-dependent mixture specific heats, $c_v = c_p - R_{mix}$.

The energy balance accounts for the enthalpy carried by every stream crossing the envelope, the leakage outflow and the moisture source included:

\begin{equation}
\begin{aligned}
c_v m \frac{dT_{room}}{dt} = {}& -c_v \left(\frac{dm_{dry}}{dt} + \frac{dm_{wv}}{dt}\right) T_{room} + \dot m_{in} c_{p,in} T_{in} \\
& - (\dot m_{out} + \dot m_{leak})\, c_p T_{room} + \dot m_{source}\, c_{p,wv} T_{source} + \dot Q
\end{aligned}
\label{eq:room-heat}
\end{equation}

where $T_{source}$ is the temperature at which the ingress vapor enters, taken as ambient.\footnote{The enthalpy of the ingress stream is carried for thermodynamic consistency. At the calibrated ingress rates its steady-state effect on the chamber temperature is below \SI{0.03}{\kelvin}, and none of the comparisons of Section 5 depend on it.}

Pressure closes pneumatically: the room acts as a capacitance whose pressure follows from its air inventory, and the outward flows follow from the pressure differences,

\begin{equation}
\begin{aligned}
P_{room} &= \frac{m_{total} R_{mix} T_{room}}{V_{room}}, \\
\dot m_{leak} &= K_{leak} (P_{room} - P_{amb}), \\
\dot m_{out} &= K_{return} (P_{room} - P_{return})
\end{aligned}
\label{eq:room-pneumatic}
\end{equation}

with $K_{return} \gg K_{leak}$ since the return duct is the intended flow path. The ratio fixes the bleed fraction by design, $\dot m_{leak}/\dot m_{RA} = K_{leak}/K_{return}$, which the measured airflow closure of the case study reproduces. One interface rule follows from the framework rather than the physics: the chamber's bus output carries ambient pressure rather than $P_{room}$, because propagating the room's gauge pressure around a recirculating topology would accumulate pressure pass over pass, an instance of the interface discipline of Section 2.1 applied to closed loops.

\subsection{Other components}

The chilled-water cooling coil is built on the Simscape Moist Air and Thermal Liquid domains, with the coil itself a thermal-liquid-to-moist-air heat exchanger \cite{mathworks2024HeatExchangerTLMA}; its governing equations are given in the cited documentation. The one point of physical consequence for this study is that the coil's outlet humidity is bounded by saturation at the coil surface temperature (the apparatus dew point), which is what limits how far condensing cooling alone can dehumidify in the bypass regime, the OFF configuration of Section 3.1 in which the recirculating air routes around the wheel. The outdoor-air pre-cooler, a packaged unit whose internal construction is not documented (Section 3), is represented by the same coil block with an equivalent geometry sized so that the coil reproduces the unit's regulated outlet temperature; its sizing is thus fixed by an independently known equipment setting rather than by any validation measurement. The remaining components (heater, humidifier, mixer, and splitter) are lightweight Simulink implementations of their elementary physics (sensible heating, vapor addition, adiabatic mixing, and flow division on the shared property functions) and warrant no further description.

\subsection{Validation methodology}

\paragraph{System-level validation} The desiccant wheel's operating point is determined jointly with the loop it serves: in recirculation, the wheel's inlet state is a function of its own past output, and the chamber, coils, and mixing all shift the equilibrium the wheel settles into. Component-level tests against isolated inlet conditions therefore cannot certify the quantity that matters, the closed-loop equilibrium, and validation in this work is performed at system level, on the generated model of the full topology, with an isolated-rotor case retained as one comparison among several rather than as the validation itself.

\paragraph{Sorbent screening and base selection} Calibration of the sorbent starts from the fact established in Section 3.3: the wheel's material is unknown, and asserting any particular library material would be a guess. All eight registered sorbents are therefore screened in the generated model, each with its isotherm taken from published measurements: zeolites 4A and 3A \cite{wang2020MeasurementsModeling}, the commercial CECA 3A, regular-density silica gel, and AQSOA Z02 \cite{goldsworthy2014MeasurementsWater}, zeolite 13X \cite{son.etal2019EquilibriumAdsorption}, MOF-801 \cite{han.chakraborty2020AdsorptionCharacteristics,furukawa.etal2014WaterAdsorption}, and AQSOA Z01 \cite{teo.etal2017ImprovedAdsorption}. Each variant is rebuilt with only the sorbent changed and run to closed-loop equilibrium under the measured operating conditions.

Screening proceeds in two stages, of which the first tests the data rather than the material. A tabulated isotherm can be evaluated only inside the temperature range its published fit covers; beyond that range the lookup holds at its boundary, so a candidate whose fit stops short of the reactivation temperature is not being simulated at the reactivation condition at all, and its outcome is not comparable with the rest. The framework raises exactly this at build time, as one of the admissibility checks of Section 2.3: a sorbent whose validity range does not span the declared regeneration temperature is flagged before the model is run. At the $129.9\,^{\circ}$C reactivation temperature delivered in this system three of the eight candidates fail the test, their published fits reaching only 100, 80, and $60\,^{\circ}$C; they are reported with the screening but excluded from the ranking. The five admissible candidates are then ranked on the process-outlet frost point their loops settle at, and the best of them, zeolite 4A \cite{wang2020MeasurementsModeling,gorbach.etal2004MeasurementModeling}, is adopted as the base shape (Section 5.2). Every subsequent adjustment acts on the one-parameter family of Eq.~\ref{eq:rotor-escale} built on that base, never on free isotherm coefficients.

\paragraph{Parameterization discipline} A system model of this scale takes dozens of inputs, and a reader is entitled to ask how many of them were adjusted until the answers came out right: with enough free parameters, agreement demonstrates nothing \cite{chong.etal2021CalibratingBuilding}. The answer is kept auditable by classifying every model input by its source (Table~\ref{tab:valid-parameterization}): physical constants and sorbent properties from the literature, geometry and ratings from datasheets and drawings, operating inputs from the measured windows, and exactly four calibrated quantities. The table states, for each calibrated quantity, the single measured constraint it was set against; the three closed-loop humidity nodes are consumed, one each, by the isotherm-family parameter and the two ingress rates, and every quantity Section 5 compares against measurement lies outside this ledger. Agreement is assessed on dew-point temperature (Section 3.3), with model states converted through the framework's own property functions so that model readout and sensor convention share the same lens. Below $0\,^{\circ}$C the facility's transmitters report the frost point (the saturation temperature over ice) per the manufacturer's specification \cite{suto2026S220Manual}, and model states are accordingly converted through the over-ice saturation branch; sub-zero agreement in Section 5 is assessed on frost point throughout.

\begin{table*}[htbp]
\centering
\caption{Model parameterization. (a) Sources of the model inputs. (b) Calibration ledger: the four calibrated quantities, each with the single measured constraint it was set against and the residual that remains.}
\label{tab:valid-parameterization}
\footnotesize
\noindent{\textbf{(a) Sources of the model inputs}}
\par\nobreak\smallskip
\begin{tabularx}{\textwidth}{@{}>{\raggedright\arraybackslash}p{0.17\textwidth} L >{\raggedright\arraybackslash}p{0.20\textwidth}@{}}
\toprule
Category & Model inputs & Source \\
\midrule
Physical constants and sorbent properties & TSL isotherm parameters (three sites: $n_s$, $b_0$, $E$) of the registered sorbents; heat of adsorption $h_{ads}$; moist-air property functions & Literature \cite{wang2020MeasurementsModeling} \\
\addlinespace
Geometry and equipment ratings & Rotor length and sector split $f_{ab} = 2/3$; channel width $w_c$; AHU coil geometry (tube count, rows, fin pitch: original drawing values); pre-cooler equivalent-coil sizing (fixed by its regulated $8\,^{\circ}$C outlet, Section 3); heater ratings; rotation speed & Datasheets, drawings, and equipment settings (Section 3) \\
\addlinespace
Fixed assumption & Solid-side LDF coefficient $k_\mathrm{LDF} = 0.04$ \si{\cubic\metre\per\kilogram\per\second} (Section 4.1) & Held fixed (Section 5.2) \\
\addlinespace
Measured operating inputs & Reference airflow structure OA 232 / EA 171 / MA 421 / SA 250 / RA 189 m$^3$/h with 61 m$^3$/h bleed (mass-balance closure of the PLC readings); boundary air states over the measurement windows; chilled-water supply temperature; regeneration air temperature at the wheel inlet, $129.9\,^{\circ}$C (panel Pt100) & Measurement windows (Sections 3.2, 4.4) \\
\bottomrule
\end{tabularx}
\par\medskip
\noindent{\textbf{(b) Calibration ledger: the four calibrated quantities}}
\par\nobreak\smallskip
\begin{tabularx}{\textwidth}{@{}>{\raggedright\arraybackslash}p{0.13\textwidth} >{\raggedright\arraybackslash}p{0.17\textwidth} >{\raggedright\arraybackslash}p{0.13\textwidth} L >{\raggedright\arraybackslash}p{0.12\textwidth}@{}}
\toprule
Symbol & Quantity & Value & Measured constraint it was set against & Residual \\
\midrule
$D_r$ & Rotor diameter & 0.65 m & Process-outlet frost point, $-40.02\,^{\circ}$C, in open-loop single pass & --- (open loop) \\
$s$ & Isotherm-family scale & 1.8 & Closed-loop equilibrium level at the process outlet & $-0.53\,^{\circ}$C \\
$\dot m_{source,AHU}$ & AHU-path moisture ingress & $1.06\times10^{-5}$ kg/s & Measured supply-air humidity gap & $-0.16\,^{\circ}$C \\
$\dot m_{source,ch}$ & Chamber moisture ingress & $2.45\times10^{-6}$ kg/s & Measured return-air humidity gap & $-0.19\,^{\circ}$C \\
\bottomrule
\end{tabularx}
\par\smallskip
\noindent\parbox{\textwidth}{\raggedright Ledger. The three closed-loop humidity nodes are consumed, one each, by $s$ and the two ingress rates; $D_r$ is set outside the loop, so no closed-loop residual attaches to it. What Section 5 then compares against measurement was available to none of the four.}
\end{table*}

\paragraph{Separation of calibration roles} Even four parameters invite a sharper objection: if a knob was turned until the model met a measurement, then meeting that measurement proves little. The calibration therefore keeps its ledger disjoint. The transport parameter $D_r$ is set by open-loop, single-pass behavior at measured inlet states, without reference to any closed-loop quantity. The isotherm-family parameter $s$ is set by the level at which the recirculating loop settles at the process-outlet station, and the two ingress rates follow in one step from the measured supply- and return-air humidity gaps: computed rather than searched, but computed from the very nodes they reconcile, so they are counted as calibration, not as evidence. Three closed-loop nodes stand against three degrees of freedom, yet none of the three knobs can drive its own residual to zero: each enters the model through a fixed physical form, a one-parameter isotherm family or a constant ingress rate, not as a per-node offset that could absorb whatever error remains. The residuals that survive are therefore informative, and their sub-degree size (Section 5) measures how well those fixed forms describe the plant. The predictive weight of the validation rests entirely on quantities this ledger never saw: the bypass regime, the reactivation power, and the measured input-step responses of Section 5.

\paragraph{Identifiability} It is fair to ask whether observations at the system's ordinary operating points could have determined the rotor parameters directly, in the manner of grey-box parameter estimation \cite{deconinck.etal2015ToolboxDevelopment}. They could not, and showing this is part of the methodology: a sensitivity analysis over fan-varied operating points, examined in the manner of practical identifiability analysis \cite{brun.etal2001PracticalIdentifiability}, shows that these data constrain a single combination of the rotor parameters, the solid-side uptake conductance, and cannot separate the desorption side from it; what does expose the desorption side is the deep-dry equilibrium level itself. Section 5.2 presents the analysis and Section 5.3 the resolution; the calibration accordingly assigns $s$ to the deep-dry closed-loop level and leaves the fan-varied data to constrain the conductance they actually measure. The methodology reports this identifiability structure rather than asserting that all parameters were identified.

\paragraph{Equilibrium qualification} Closed-loop model results are admitted into the validation only when qualified by drift rate over the terminal evaluation window (drift below $0.0040\,^{\circ}$C per 1000 s), not by their endpoint value; slow drifts masquerade as plausible endpoints long before equilibrium, and endpoint-based comparison would reward them.

\paragraph{Energy comparison} Over the same measurement windows as the dew-point comparisons, per-device electric power channels are averaged and compared with the corresponding model duties. The primary energy comparison is the reactivation heater, a resistive load for which electrical power equals thermal duty with minimal assumption. Interpretation of the power channels is part of the methodology: the heater's metering covers one of its two heating banks, established by circuit inspection, by the ratio of logged power to duty over sixty hours of operation, and by consistency of the back-calculated reactivation inlet temperature; the comparison accordingly uses twice the metered value. Standby consumption of the desiccant unit in the bypass regime lies outside the model's scope and is excluded from the comparison.

\section{Results}\label{sec:results}

\subsection{Generation efficiency (C1)}

The framework's first claim is that a validated physical model of an HVAC system can be produced from a declarative specification at a small fraction of the effort of building it by hand. We measured this on the case-study topology (topo14, the dehumidification configuration) by comparing an expert building the model manually in Simulink against the generator, both starting from the same component library and the same parameter sheet, so that the comparison isolates construction effort rather than the parameter research that precedes it (Table~\ref{tab:results-time-study}).

\begin{table}[htbp]
\centering
\caption{Construction effort for the same validated topology: expert manual build versus automated generation.}
\label{tab:results-time-study}
\begin{tabularx}{\textwidth}{@{}l L L@{}}
\toprule
Metric & Manual (expert) & Automated \\
\midrule
Time to runnable model & 22 min 12 s (hands-on) & \textbf{87.9 s} (cold, engine start-up incl.) / $\sim$44 s (warm) \\
Errors / rework & 1 (missing solver max-step $\rightarrow$ oscillation $\rightarrow$ 1 min 40 s rework) & 0 (3/3 deterministic) \\
Parameter entries & 47 + init 4 + solver 3 (manual) & 0 (from YAML) \\
Wiring operations & 28+ (manual) & 0 \\
Reproducibility & --- & bit-for-bit identical (3 runs) \\
\textbf{Speed-up} & --- & \textbf{15$\times$ (cold, conservative) / $\sim$30$\times$ (warm)} \\
\bottomrule
\end{tabularx}
\end{table}

Manual construction reached a runnable model in 22 min 12 s of hands-on work and incurred one rework cycle, a missing solver max-step setting that produced numerical oscillation and had to be diagnosed and corrected; it required 47 mask-parameter entries and on the order of 28 wiring operations entered by hand. The generator produced the same model in 87.9 s from a cold start, engine start-up included, and in roughly 44 s warm, with no manual parameter entry and no wiring: every value and every connection derives from the YAML file. The construction is deterministic: regenerating from the same specification reproduces the model identically, and three successive generations were bit-for-bit identical. Taken conservatively against the cold-start time, this is a fifteenfold reduction in construction effort, rising to about thirtyfold warm.

\begin{figure}[htbp]
\centering
\includegraphics[width=\textwidth]{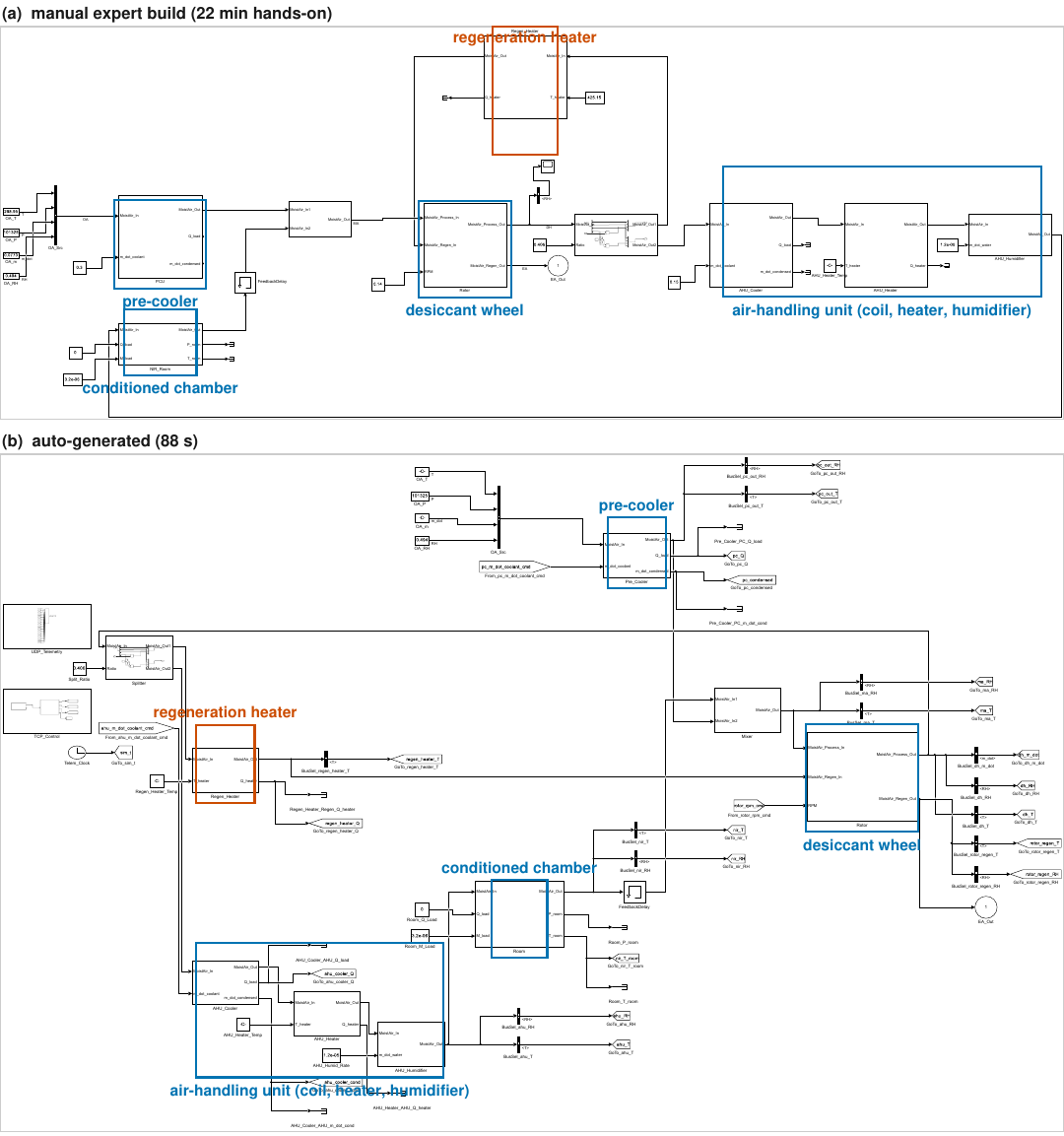}
\caption{Top-level diagrams of the same topology built (a) manually by an expert user (22 min hands-on, one rework cycle) and (b) by the generator (88 s including engine start-up): connectivity is identical, while the generated layout is deterministic and reproducible. Colored frames mark the principal components in both panels (regeneration side in orange).}
\label{fig:results-manual-vs-auto}
\end{figure}

Two qualifications attach to the number. The comparison excludes the communication and telemetry layer, the taps that make the model twin-ready, because building those by hand is not something an expert completes in a comparable sitting; the measured ratio is therefore a lower bound on the automation gain, taken over only the part a human can realistically do. And the value is not the speed as such but that the declarative route yields a model carrying the physical depth documented in Section~\ref{sec:validation}; the fifteenfold figure is the token of that route rather than its point. Figure~\ref{fig:results-manual-vs-auto} places the two top-level diagrams side by side: the connectivity is identical, and the layouts differ only in the incidental choices a human makes and a generator does not. That they can differ at all without either being wrong is the point: once the wiring is complete and correct, the arrangement on the canvas is a convenience for the human reader, not a property of the model.

\subsection{The identification problem}

With the construction claim established, the remainder of this section concerns the physical validation. It begins not with agreement but with the obstacle that has to be cleared before agreement means anything.

\begin{figure}[htbp]
\centering
\includegraphics[width=\textwidth]{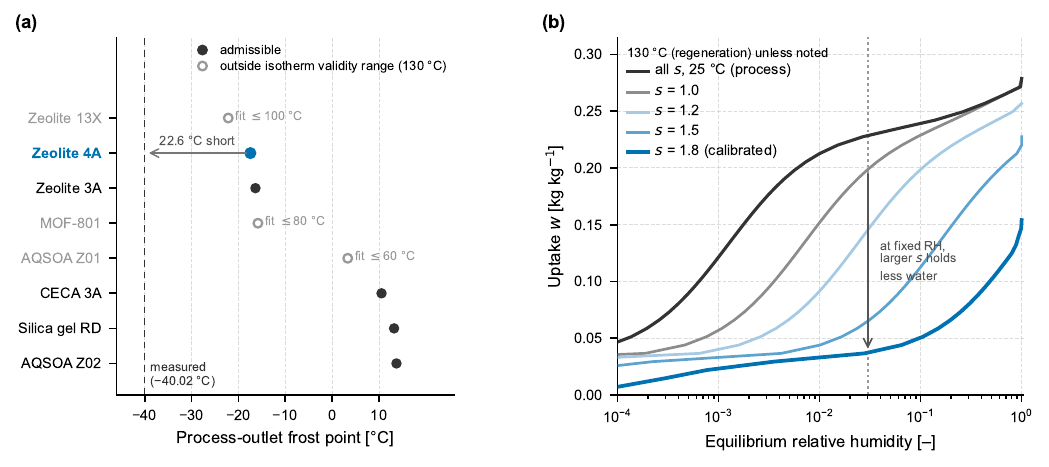}
\caption{Sorbent-side calibration. (a) Two-stage screening of the eight registered sorbents at the measured operating condition: candidates whose published isotherm fit does not span the $129.9\,^{\circ}$C reactivation temperature are drawn open and excluded from the ranking (Section 4.4), and of the five admissible candidates none reaches the measured process-outlet frost point, the best falling $22.6\,^{\circ}$C short. Sub-zero values are frost points (Section 4.4). (b) The E-scaled isotherm family: renormalizing the site-affinity pre-exponentials pins the process-temperature isotherm (the $25\,^{\circ}$C curve is common to all variants) while amplifying only its decline toward regeneration temperature, so the scale factor $s$ acts as a pure desorption-side knob; $s = 1.8$ is the calibrated member.}
\label{fig:results-isotherm-escale}
\end{figure}

The sorbent of the wheel is undisclosed (Section~3.3), so no isotherm can be asserted a priori, and the two-stage screening of Section~4.4 was run over all eight registered materials (Fig.~\ref{fig:results-isotherm-escale}a). Two facts follow.

The first is that no library isotherm reaches the measured depth unmodified. The admissible candidates settle between $-17.5$ and $+13.7\,^{\circ}$C at the process outlet against $-40.02\,^{\circ}$C measured, the best of them short by $22.6\,^{\circ}$C, and no candidate at all, the three excluded ones included, comes within $17\,^{\circ}$C of the measurement. Some calibration of the isotherm is therefore unavoidable, and the question is only what is allowed to move. One of the excluded candidates, zeolite 13X, does settle deeper than any admissible one; it is set aside because its published fit ends at $100\,^{\circ}$C and so cannot be evaluated at this system's reactivation temperature, not because it performs poorly, and since boundary clamping understates desorption its true depth would be greater still. Admitting it would mean extrapolating a fitted isotherm $30\,^{\circ}$C beyond its measured range, which is the kind of unstated assumption the screening is meant to keep out.

The second is what separates the candidates, and it is not the material. Among the five admissible the split is complete: the two triple-site Langmuir zeolite fits settle on the dry branch and all three fits carried in Hill form stay near saturation, with $27\,^{\circ}$C of empty range between the groups. Yet the CECA 3A fit is nominally the same zeolite as candidate 3A, and carries the same isotherm classification, and it stays trapped. What the screening separates is the isotherm fit and its formulation rather than the nominal material class, which is the same reason the wheel's own sorbent cannot simply be looked up. Zeolite 4A, the best of the admissible candidates and marginally ahead of zeolite 3A, is adopted as the base shape on which the one-parameter family of Section~4.1 is built.

The screening, like the parameter perturbations below and the isotherm-family sweep of Section~5.3, consists of generated model variants, each rebuilt from a lightly edited specification; a calibration of this breadth is practical only because construction is cheap, which is where the efficiency of Section~5.1 stops being a convenience and becomes part of the validation methodology.

\begin{figure}[htbp]
\centering
% width 0.90: at full \linewidth the float with this caption exceeds the text height
\includegraphics[width=0.90\linewidth]{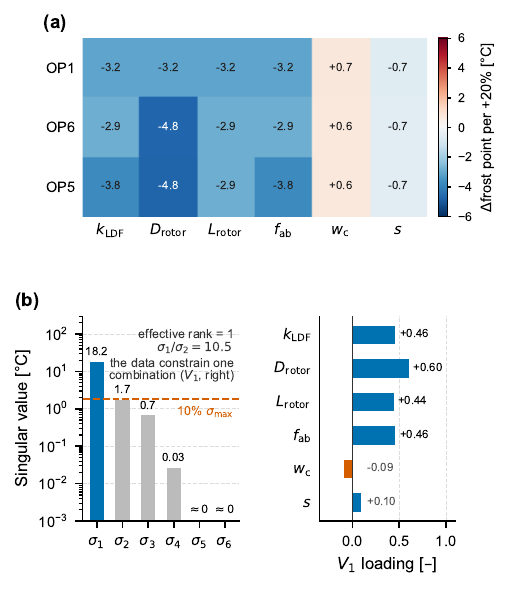}
\caption{Identifiability of the wheel parameters at the system's ordinary operating points. (a) Chamber frost-point sensitivity to +20\% parameter perturbations over fan-varied operating points (heatmap): the kinetic coefficient, rotor dimensions, and sector fraction respond in near-fixed proportion, while the channel width and the isotherm-family scale $s$ respond at a fraction of their size. (b) Singular-value spectrum of the chamber and supply-air sensitivity matrix: effective rank one ($\sigma_1/\sigma_2 \approx 10$), with the dominant direction $V_1$ identifying the solid-side uptake conductance $k_{LDF}\, m_r$ (with $m_r \propto D_r^2 L_r f_{ab}$); the response to $s$ lies in the near-null space, so the desorption side is unidentifiable from these data (Section~5.2).}
\label{fig:results-identifiability}
\end{figure}

Granting that a calibration is needed, the sharper question is whether observations at the system's ordinary operating points can determine the wheel's parameters individually, or only some combination of them. They determine only a combination. Perturbing each rotor parameter by +20\% (the kinetic coefficient $k_{LDF}$, the dimensions $D_r$ and $L_r$, the sector fraction $f_{ab}$, the channel width $w_c$, and the isotherm-family scale $s$) and recording the chamber and supply-air dew-point responses over fan-varied operating points yields the sensitivity matrix

\begin{equation}
S_{ij} = \frac{\partial T_{dp,i}}{\partial \ln \theta_j} \approx \frac{\Delta T_{dp,ij}}{0.20}
\label{eq:sensitivity}
\end{equation}

whose rows are the measurement stations at the three fan-varied operating points, whose columns are the six wheel parameters $\theta_j$, and whose entries carry units of $^{\circ}$C (Table~\ref{tab:results-sensitivity}). The kinetic coefficient, rotor length, and sector fraction respond at $-2.9$ to $-3.8\,^{\circ}$C per perturbation, the diameter at $-3.2$ to $-4.8\,^{\circ}$C with its excess over the others varying by operating point, the channel width at a fraction of a degree, and $s$ at under a degree. The singular-value spectrum of the matrix has an effective rank of one: the second singular value sits an order of magnitude below the first ($\sigma_1/\sigma_2 \approx 10$), with dominant direction $V_1 \propto [1,\ 1.3,\ 1,\ 1,\ {\approx}0,\ {\approx}0]$ over these six axes (Fig.~\ref{fig:results-identifiability}). In physical terms, each singular value is the frost-point response, in $^{\circ}$C, that the measurements offer along one independent combination of fractional parameter changes, and the entries of $V_1$ are the weights of the best-observed combination: these operating points respond by about $18\,^{\circ}$C along the direction that moves $k_{LDF}$, $L_r$, $f_{ab}$, and (with weight 1.3) $D_r$ together, and by less than $2\,^{\circ}$C along every direction orthogonal to it.

\begin{table}[htbp]
\centering
\caption{Chamber frost-point sensitivity to +20\% parameter perturbations over fan-varied operating points, and the singular-value structure of the chamber and supply-air sensitivity matrix.}
\label{tab:results-sensitivity}
\begin{tabularx}{\textwidth}{@{}l L L@{}}
\toprule
Parameter & $\Delta$frost point per +20\% ($^{\circ}$C, chamber) & Note \\
\midrule
$k_{LDF}$ (LDF coefficient) & $-2.9$ to $-3.8$ & product-direction component 1 \\
$D_r$ (rotor diameter) & $-3.2$ to $-4.8$ & weight $\approx 1.3$, operating-point dependent; the excess flow-side effect forms the second direction \\
$L_r$ (rotor length) & $-2.9$ to $-3.2$ & component $\approx 1$ \\
$f_{ab}$ (adsorption-sector fraction) & $-2.9$ to $-3.8$ & component 1 \\
$w_c$ (channel width) & $\approx +0.6$ & $\approx 0$ (outside the collinear subspace) \\
$s$ (isotherm E-scale) & $-0.66$ to $-0.70$ (supply air $-0.75$ to $-0.82$) & weak and degenerate: parallel to the conductance direction, carried by $V_4$ ($\sigma_4/\sigma_1 \approx 0.0015$) \\
\midrule
\multicolumn{3}{@{}>{\raggedright\arraybackslash}p{\textwidth}@{}}{\textbf{SVD} (chamber and supply air, $6 \times 6$): $\sigma = [18.2,\ 1.7,\ 0.7,\ 0.03,\ {\sim}0,\ {\sim}0]$, $\sigma_1/\sigma_2 = 10.5$, effective rank 1; $V_1 \propto [1,\ 1.3,\ 1,\ 1,\ {\approx}0,\ {\approx}0] \rightarrow k_{LDF}\, m_r$} \\
\bottomrule
\end{tabularx}
\end{table}

That direction has a precise physical reading. The matrix inventory scales as $D_r^2 L_r f_{ab}$, so what the steady-state data constrain is essentially the single product $k_{LDF}\, m_r$, the solid-side uptake conductance; the diameter's additional flow-side effect supplies the weak second direction. Two consequences follow. Any combination of parameters preserving the conductance product is indistinguishable at these operating points. And the desorption side, though not strictly invisible, is unidentifiable here: the response to $s$ is a fifth the size of the conductance responses and nearly parallel to them, so no fit to these data can separate a change in the isotherm family from a small change in conductance. (The wheel's own outlet station is excluded from the matrix: as its air approaches zero humidity the frost-point scale stretches without bound, and temperature-unit sensitivities there are ill-posed; the next section turns that same stretching into the identifying signal.) A steady-state fit at ordinary operating points therefore determines one number, and the regeneration side of the isotherm must be identified elsewhere.

\subsection{Resolution by the deep-dry closed-loop level}

\begin{figure}[htbp]
\centering
\includegraphics[width=\linewidth]{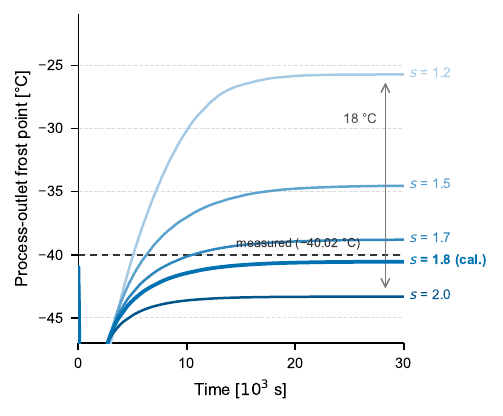}
\caption{Level-based identification of the isotherm family in recirculation. Closed-loop process-outlet frost-point trajectories for the variants $s = 1.2$ to 2.0 at the measured regeneration boundary; the dashed line is the measured level and the calibrated member $s = 1.8$ is drawn in bold. Each variant settles at a level of its own, and it is that level, not the fact of settling, which identifies $s$ (Section~5.3).}
\label{fig:results-closedloop-levels}
\end{figure}

The regeneration side is identified where its effect is large. The isotherm-family scale $s$ is pinned at process temperature by construction (Section~4.1), and at the fan-varied operating points of Section~5.2 its trace is marginal; as the system dries, the wheel's working state couples ever more strongly to how completely the matrix regenerates, and the response to $s$ grows by an order of magnitude. In single-pass operation at the measured deep-dry inlet, the variants $s = 1.5,\ 1.7,\ 2.0$ place the process-outlet frost point at $-31.6$, $-35.7$, and $-40.1\,^{\circ}$C, a spread of $8.5\,^{\circ}$C between isotherms that the fan-varied observations of Section~5.2 could barely tell apart. In recirculation the same separation appears as the level at which the loop settles. At the measured regeneration boundary, every variant settles into a stationary equilibrium of its own (terminal drift rates at or below $0.0083\,^{\circ}$C per 1000 s): the grid $s = 1.2$--$2.0$ spans about $18\,^{\circ}$C at the process outlet, a step of 0.1 in $s$ near the measured level moves the equilibrium by about $1.7\,^{\circ}$C, and the calibrated $s = 1.8$ settles within $0.6\,^{\circ}$C of the measured level (Fig.~\ref{fig:results-closedloop-levels}).

This deep-dry equilibrium level is the measurement the calibration assigns to $s$ (Section~4.4). The assignment is bookkeeping, not a prediction (the level is a fitted constraint and is counted as one), but it settles where the information lives: the desorption side of an effective isotherm is pinned by the deep-dry equilibrium the loop itself reaches, not by accumulating further steady-state points at ordinary operating conditions, which Section~5.2 shows cannot separate it from the transport parameters. For recirculating dehumidification systems this identifies which measurement a generated twin actually needs, and it is why the validation that follows rests on quantities withheld from the calibration rather than on the level that identified $s$.

\subsection{Prediction against withheld measurements}

\begin{figure}[htbp]
\centering
\includegraphics[width=\textwidth]{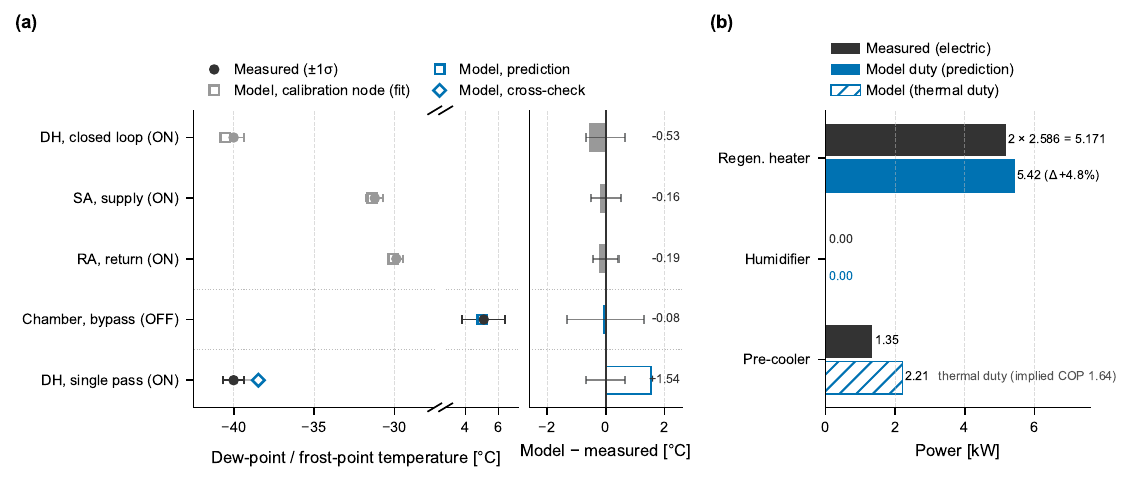}
\caption{Model--measurement agreement across both operating regimes with a single parameter set. (a) Humidity comparison: the three closed-loop nodes consumed by calibration (muted, shown as fit consistency), the bypass-regime chamber (prediction), and the isolated single-pass cross-check; sub-zero values are frost points (Section~4.4). Filled markers show measurement-window means with $\pm 1\sigma$ bands, open markers the drift-qualified model equilibria; the right panel gives the residuals. (b) Electric-power comparison over the same windows: the regeneration heater (twin-bank metering, $2 \times 2.586 = 5.171$ kW, prediction) and the humidifier; pre-cooler electric consumption is reported alongside the model thermal duty (implied COP, not directly comparable).}
\label{fig:results-anchors}
\end{figure}

The calibration of Section~4.4 consumes the three closed-loop humidity nodes and nothing else. With the single parameter set fixed (Zeolite 4A E-scaled at $s = 1.8$, $D_r = 0.65$ m, $f_{ab} = 0.667$, and the two ingress rates), the generated closed-loop model settles within $0.6\,^{\circ}$C of all three: process-outlet frost point $-40.55\,^{\circ}$C against $-40.02$ measured, supply air $-31.39$ against $-31.23$, return air $-30.08$ against $-29.89$, every equilibrium drift-qualified in the sense of Section~4.4. These residuals measure the consistency of the calibration's fixed physical forms, and this paper does not count them as predictions. The predictions are what the same parameter set gets right about measurements it never saw (Table~\ref{tab:results-predictions}, Fig.~\ref{fig:results-anchors}). In the figure, each measured value is the mean of the logged samples over its steady evaluation window, with an error bar spanning one standard deviation of those samples; model values are single drift-qualified equilibria (Section~4.4) and carry no comparable statistic.

\begin{table}[htbp]
\centering
\caption{Prediction against measurements withheld from calibration: the bypass-regime chamber, the reactivation-heater power, and three measured input steps (rotor speed, chilled water, air-handling heater setpoint), each compared with the response of the same calibrated parameter set, including the two cases where the correct response is none.}
\label{tab:results-predictions}
\begin{tabularx}{\textwidth}{@{}L r >{\raggedright\arraybackslash}p{3.6cm} l@{}}
\toprule
Quantity (withheld from calibration) & Model & Measured & $\Delta$ \\
\midrule
Bypass-regime chamber dew point ($^{\circ}$C) & $+5.02$ & $+5.10$ & $-0.08$ \\
Reactivation heater duty (kW) & 5.42 & 5.171 ($2 \times 2.586$, twin-bank) & $+4.8$\% \\
Rotor 60 to 50 Hz: chamber dew-point response ($^{\circ}$C) & $-0.4$ & $\leq 0.1$ (below noise) & null-match \\
Chilled water 5 to $7\,^{\circ}$C: supply-air dew-point response ($^{\circ}$C) & $\leq 0.06$ & $\pm 0.5$--$0.9$ (noise wander) & null-match \\
AHU heater 20 to $22\,^{\circ}$C: supply-air temperature response ($^{\circ}$C) & $+2.19$ & $+2.0$ & $+0.19$ \\
\bottomrule
\end{tabularx}
\end{table}

The first is the other operating regime. In bypass mode (a different generated topology, the wheel out of the path, the chamber held by the condensing coil alone) the model settles at a chamber dew point of $+5.02\,^{\circ}$C against $+5.10$ measured. The second is the energy channel: the reactivation heater, never a calibration target, is metered at 5.171 kW against a model reactivation duty of 5.42 kW, an agreement of +4.8\% obtained with the regeneration inlet temperature held at its measured value rather than adjusted (metering interpretation in Section~5.5).

The remaining rows of the table are measured input steps, each compared within its own measurement session so that the comparison is a response, not a level. Where the plant does not respond, the model predicts that it should not: a rotor-speed step from 60 to 50 Hz moves the measured chamber dew point by no more than $0.1\,^{\circ}$C, and the model responds with $-0.4\,^{\circ}$C at the chamber ($-1.2\,^{\circ}$C at the process outlet, where the measured channel itself wanders non-monotonically by about $1\,^{\circ}$C); a chilled-water step from 5 to $7\,^{\circ}$C leaves the measured supply-air dew point inside its $\pm 0.5$--$0.9\,^{\circ}$C wander, and the model responds by no more than $0.06\,^{\circ}$C, the coil being non-condensing at these states. Where the plant does respond, the model tracks it: an air-handling heater setpoint step from 20 to $22\,^{\circ}$C raises the measured supply-air temperature by $2.0\,^{\circ}$C, against $2.19\,^{\circ}$C in the model. The humidity channel of that step is excluded from the comparison: the plant's humidification is governed by a relative-humidity controller, so a temperature step moves its humidity command through the control layer, which the physical model does not represent. Finally, an isolated single-pass model of the calibrated wheel, driven by measured boundary conditions, reproduces the process-outlet frost point to $+1.54\,^{\circ}$C; this is reported as a cross-check rather than a prediction, since its mixed-air inlet state is itself an estimate.

\subsection{Energy comparison}

Over the same measurement windows as the humidity comparisons, the per-device electric-power channels are compared with the corresponding model duties (Table~\ref{tab:results-predictions}, Fig.~\ref{fig:results-anchors}). The primary comparison is the reactivation heater, the dominant energy input of desiccant dehumidification \cite{abdelgaied.etal2019PerformanceImprovement} and a resistive load whose electrical power equals its thermal duty with minimal assumption. The heater is built as two identical banks and the facility meters one of them, so the comparison uses twice the metered value: $2 \times 2.586 = 5.171$ kW against a model reactivation duty of 5.42 kW, an agreement of +4.8\%, obtained with the regeneration inlet temperature held at its measured value (Section~4.4) rather than adjusted.

A second, qualitative check is the humidifier, metered at zero throughout, consistent with its being off in both regimes. The pre-cooler is reported as corroboration rather than a calibrated match: its metered electric power set against the model's thermal duty implies a coefficient of performance of about 1.6, plausible for the small packaged refrigeration unit it is, though the package exposes no thermal measurement that would allow a closer comparison (Section~3.2). The cold source of the air-handling coil (a precision chiller that runs its compressor continuously and trims with a heater) does not admit an electric-to-duty comparison at all, and its channels, together with the fans and the desiccant unit's standby consumption in the bypass regime, lie outside the comparison.

\section{Discussion}\label{sec:discussion}

What an automated generation framework finally owes its user is not a model that runs but a model that behaves like the equipment. The executable model of Section~2 is the necessary half of that deliverable; the calibrated, measurement-tested twin of Sections~4 and~5 is the other, and this is why a paper about automatic construction spends much of its length on calibration and validation: the two are stages of one pipeline. The dependence also runs in the practical direction, as Section~5.2 noted: every screening, sweep, and perturbation of the calibration is a regenerated model variant, an experimental style that hand-built models do not support at reasonable cost. Generation without validation is unfinished, and validation of this breadth is impractical without generation.

The calibration that closes the pipeline must then withstand the obvious objection: with enough adjustment, any model meets any measurement. The answer is structural rather than arithmetic, and it is kept in one place: the ledger of Section~4.4 names each of the four calibrated quantities and the single measurement it was set against, the three closed-loop humidity nodes are thereby spent, and their sub-degree residuals are counted as consistency of the fixed physical forms, not as evidence. What the model is judged on is everything the ledger never touched: the bypass regime, the reactivation power, the measured input-step responses, reproduced by the same parameter set with no further adjustment (Section~5.4).

Nor are the freedoms interchangeable. Each calibrated quantity draws on a different aspect of the data (single-pass depth, closed-loop level, station-to-station humidity gaps), and each enters through a fixed physical form with no freedom to absorb the others' errors; in particular, the identifiability analysis of Section~5.2 is why the isotherm-family parameter could not have been set at ordinary operating points and is set instead at the deep-dry level, where its effect is an order of magnitude larger.

The choice of knob itself rests on mechanism rather than search. Deep dehumidification lives or dies on the desorption side: each pass, the wheel can remove only what regeneration has cleared from the matrix, so the depth the recirculating loop reaches is governed by how completely the isotherm declines toward regeneration temperature, the dependence Section~5.3 exposes as the separation of the family variants' equilibrium levels. The scale factor $s$ is the parameter this mechanism points to, not the parameter a residual search happened to land on.

What $s = 1.8$ means is then worth stating plainly. Read as a correction to zeolite 4A itself it would nearly double the site activation energies (58.6 to 105.5 kJ/mol), well beyond the measured sorption energetics of water on zeolite 4A \cite{gorbach.etal2004MeasurementModeling}. But the calibration claims no such thing, because the wheel's sorbent is not zeolite 4A: it is an undisclosed commercial formulation for which 4A is only the screened base shape (Section~4.4), and the family parameter was constructed to carry exactly the one part of the isotherm that no process-condition measurement of any candidate material could pin: its decline toward regeneration temperature. What the calibrated value asserts is that the actual matrix retains substantially less moisture at 130\,$^{\circ}$C than the 4A literature isotherm would predict; operationally, that its regeneration is nearly complete at rated reactivation temperature, which is what a wheel engineered for deep dehumidification is built to achieve. Part of the value may equally be compensating desorption-side physics that the two-zone lumped closure omits (the angular distribution of carryover, gradients within a sector, the temperature dependence of the sorption kinetics), and the present data cannot apportion the excess between material difference and model form. We therefore report $s$ as an effective parameter and its magnitude as the measured size of that combined gap, not as a sorption property of anything.

Resolving the angular distribution within a sector would remove the model-form part of that ambiguity, and we did not do it. The choice is a trade-off, not an oversight: a generated twin is built to be regenerated in numbers and searched across operating points, and spatial discretization buys fidelity with exactly the currency that purpose spends, simulation speed and deployability. Our position is to let the lumped closure absorb what it can, which is what the effective isotherm family does, and to pay for spatial resolution only where the physics refuses to be absorbed; nothing in the present measurements shows that refusal.

The methodological finding beneath the calibrated value is the one Section~5 demonstrated: steady-state observation of this system at ordinary operating points constrains a single solid-side conductance product and cannot separate the desorption branch from it, and the deep-dry equilibrium level the recirculating loop itself reaches is the datum that identifies the remainder. For recirculating dehumidification systems, that is a statement about which measurements a generated twin actually needs, arguably more transferable than any parameter value in this paper.

The calibration's validity domain reads the same way. At the manufacturer's rated design point, whose inlet condition lies outside the facility's operating windows, the calibrated wheel reproduces the direction of the rated performance but falls roughly 8\,$^{\circ}$C short of its depth in frost point, even though the facility's own deep-dry outlet sits at the rated humidity (0.078 against 0.08 g/kg), so the shortfall belongs to the effective parameters, not the plant. An effective isotherm identified at one site's regimes carries those regimes; we report the design-point comparison as the measured edge of that domain rather than absorb it, because absorbing it would cost exactly the parameterization discipline of Section~4.4 that the rest of the argument stands on. A held-out flow sweep from the same measurement period, the supply fan stepped across its range within one session, localizes the one structural simplification the data exposed: with the constant ingress rates of Table~\ref{tab:valid-parameterization} the sweep's humidity spans do not close, and reconstructing the ingress from the monitored humidity gaps at each fan setting closes them while following a near-quadratic pressure law. The refinement this marks is a pressure-dependent ingress submodel, a change of boundary model, not of wheel physics.

The limitations are the mirror of these choices. The sorbent remains unidentified, so the calibrated isotherm is a surrogate fitted to measurement: adequate for the twin's purpose, but not a characterization of the material. Since undisclosed sorbents are the industrial norm for commercial wheels, we regard the surrogate procedure as part of the method rather than a compromise of it. The validation covers one site, two operating regimes, and the measured input steps of Section~5.4; a regeneration-temperature sweep and genuine transient validation were not available within the facility's duty schedule and stand as the natural next experiments, and the boundary accounting would benefit from logged outdoor-air humidity, whose absence restricted the flow-sweep comparison to within-session spans. Validation is system-level by necessity (Section~4.4): it certifies the closed-loop behavior the application cares about, but it does not certify each component in isolation. Finally, some facts about a system belong to no single block: the two cooling coils of this plant draw on separate cold sources (a precision chiller dedicated to the air-handling coil, and a packaged refrigeration unit inside the desiccant system), yet the specification can state each coil's cold side only as parameter values local to that coil, with no way to say whether two coils share a plant or, as here, do not. Identity and distinctness of resources are system-level facts, and expressing them calls for an ontology above the component library, in the spirit of building metadata schemas \cite{balaji.etal2018BrickMetadata}, which we leave as future work.

The calibration procedure, finally, is systematized but not automated: screening, single-pass calibration, level-based identification, and drift qualification each run on explicit, machine-checkable criteria over machine-readable inputs. Folding that loop into the generation pipeline, so that a specification, a component library, and a measurement window yield a calibrated twin without an analyst in between, is the natural continuation of this work, for which scalable calibration machinery is already demonstrated \cite{chakrabarty.etal2021ScalableBayesian}, and operating such twins on-line against the live system is the subject of a companion effort. The fourteen topology specifications built with the framework over the course of this project suggest the construction side generalizes; what this paper establishes is the harder half of the claim: that a generated model can be brought to answer to measurement, with its remaining freedoms counted and its remaining errors named.

\section{Conclusion}\label{sec:conclusion}

This paper set out to remove the construction of the model itself as the bottleneck of physics-based HVAC digital twins. The framework presented here generates a dynamic Simulink twin from a declarative topology specification: a purpose-built library of physical component models sharing a standard moist-air bus, compiled into a wired, solver-configured, telemetry-equipped model with no manual assembly. The physics carried by the generated models reaches the depth that low-dew-point applications require, including a desiccant-wheel model whose sorption isotherm is an interchangeable component, which is what allows a wheel of undisclosed sorbent, the ordinary condition of commercial rotors, to be calibrated as an effective isotherm under an explicit identifiability discipline.

On the industrial-grade low-dew-point system designed and built for this study, the framework reached a runnable model in 88 seconds against 22 minutes of expert manual construction, a fifteenfold reduction taken conservatively. A single calibrated parameter set, with four degrees of freedom whose ledger Section~\ref{sec:validation} states in full, holds the dehumidification loop within $0.6\,^{\circ}$C of its three measured humidity nodes and then, without further adjustment, predicts the bypass-regime chamber to within $0.1\,^{\circ}$C, the reactivation-heater power to within 5\%, and the measured responses to rotor-speed, chilled-water, and heater-setpoint steps, including the two cases where the correct response is none.

The validation also produced a finding about method that we expect to outlive this particular system. Steady-state observations of a recirculating dehumidification loop at ordinary operating points constrain a single solid-side conductance product and cannot separate the desorption side of the isotherm from it, however many such points are measured; the deep-dry equilibrium level that the loop itself reaches is the datum that identifies the remainder. For calibrating generated twins of recirculating systems, this points away from accumulating further steady-state points at moderate conditions and toward the loop's own deep equilibrium as the informative measurement.

The validation covers one site, two regimes, and a set of measured input steps, and the calibrated isotherm remains a surrogate for an unidentified material; a regeneration-temperature sweep and transient step tests are the natural next experiments. Beyond that, the calibration procedure itself (screening, single-pass calibration, level-based identification, drift qualification) runs on explicit criteria over machine-readable data, and folding it into the generation pipeline would carry automation the rest of the way from plant description to calibrated twin. Operating such twins on-line, synchronized with the live plant and searching for better operating points, is the subject of ongoing work.

%% Elsevier 필수 back matter — CRediT / 이해상충 / 감사의 글
%% CRediT 역할 배분 = 사용자 확정 (2026-08-06). 자금 = KIER 주요사업 C6-2419-63.
\section*{CRediT authorship contribution statement}

\textbf{Younghwan Joo:} Conceptualization, Methodology, Software, Validation,
Formal analysis, Investigation, Data curation, Visualization, Writing --
original draft, Writing -- review \& editing.
\textbf{Jeonghoon Han:} Investigation, Resources, Visualization, Writing --
review \& editing.
\textbf{Sang Hyun Oh:} Investigation, Resources, Writing -- review \& editing.
\textbf{Soosik Bang:} Formal analysis, Validation, Writing -- review \& editing.
\textbf{Sung-il Kim:} Supervision, Project administration, Funding acquisition,
Writing -- review \& editing.

\section*{Declaration of competing interest}

The authors declare that they have no known competing financial interests or
personal relationships that could have appeared to influence the work reported
in this paper.

\section*{Acknowledgements}

This work was conducted under the framework of Research and Development Program
of the Korea Institute of Energy Research (KIER) (C6-2419-63).

\section*{Data availability}

The measurement and validation data supporting this study are openly available
at figshare (DOI: 10.6084/m9.figshare.33170660). The archived dataset comprises the humidity and
electric-power validation measurements together with their metering-window
statistics, the closed-loop frost-point time series used for the drift-based
equilibrium assessment, the measured response to the operating levers, the
static single-pass reference measurements, and the outputs of the sensitivity
and identifiability analysis.
A data dictionary maps each measurement channel to the corresponding node in
the paper, and documents the conditioning applied to the raw channels: the
exclusion of the chamber dew-point channel for the high bias noted in
Section~4, and the two-bank interpretation of the reactivation-heater power
metering.

Equipment specifications derive from manufacturer datasheets and are not
redistributable; in particular, the desiccant sorbent of the case-study wheel
is undisclosed by the manufacturer, which is a premise of the calibration
methodology rather than an omission. The model-generation framework and the
generated Simulink models are not publicly released.

\section*{Nomenclature}

\noindent\textbf{Latin symbols}
\begin{xltabular}{\linewidth}{@{}l L l@{}}
\toprule
Symbol & Description & Unit \\
\midrule
\endhead
$A_c$ & sector face area & m$^2$ \\
$A_s$ & transfer (surface) area & m$^2$ \\
$b_k$, $b_{0,k}$ & TSL site affinity and its pre-exponential & --- \\
$C$, $C^*$ & vapor concentration; equilibrium concentration at matrix surface & kg\,m$^{-3}$ \\
$c_p$, $c_v$ & specific heat at constant pressure / volume & J\,kg$^{-1}$\,K$^{-1}$ \\
$D_h$ & hydraulic diameter & m \\
$D_r$ & rotor diameter & m \\
$E_k$ & TSL site activation energy & J\,mol$^{-1}$ \\
$f_{ab}$, $f_{zone}$ & adsorption-sector / zone area fraction & --- \\
$h$ & gas-film heat-transfer coefficient & W\,m$^{-2}$\,K$^{-1}$ \\
$h_{ads}$ & specific heat of adsorption & J\,kg$^{-1}$ \\
$K_{leak}$, $K_{return}$ & leakage / return-path flow conductance & kg\,s$^{-1}$\,Pa$^{-1}$ \\
$k_{eff}$ & effective (series) mass-transfer coefficient & m\,s$^{-1}$ \\
$k_{LDF}$ & solid-side linear-driving-force coefficient & m$^3$\,kg$^{-1}$\,s$^{-1}$ \\
$k_m$ & gas-film mass-transfer coefficient & m\,s$^{-1}$ \\
$L_r$ & rotor length & m \\
$Le$ & Lewis number & --- \\
$m_{dry}$, $m_{wv}$ & dry-air / water-vapor mass in chamber & kg \\
$m_{r,i}$ & desiccant matrix mass of zone $i$ & kg \\
$\dot m$ & mass flow rate & kg\,s$^{-1}$ \\
$\mathrm{NTU}$, $\mathrm{NTU}_m$ & number of transfer units (heat / mass) & --- \\
$Nu$ & Nusselt number & --- \\
$P$ & pressure & Pa \\
$p_{sat}$ & saturation vapor pressure & Pa \\
$\dot Q$ & heat load & W \\
$R$ & universal gas constant & J\,mol$^{-1}$\,K$^{-1}$ \\
$R_w$, $R_{mix}$ & specific gas constant (water vapor / moist-air mixture) & J\,kg$^{-1}$\,K$^{-1}$ \\
$RH_{eq}$ & equilibrium relative humidity (isotherm closure) & --- \\
$S_{ij}$ & sensitivity-matrix entry & $^{\circ}$C \\
$s$ & isotherm-family scale factor & --- \\
$T$ & temperature & K \\
$T_{dp}$ & dew-point (below $0\,^{\circ}$C, frost-point) temperature & $^{\circ}$C \\
$T_{r,i}$ & matrix temperature of zone $i$ & K \\
$T_{ref}$ & reference temperature (298.15 K) & K \\
$V$ & channel velocity & m\,s$^{-1}$ \\
$V_{room}$ & chamber volume & m$^3$ \\
$V_1$ & dominant right singular vector & --- \\
$w_c$ & channel width & m \\
$w_{r,i}$ & matrix moisture content of zone $i$ & kg\,kg$^{-1}$ \\
$Y$ & humidity ratio & kg\,kg$^{-1}$ \\
\bottomrule
\end{xltabular}

\vspace{1em}
\noindent\textbf{Greek symbols}\par\nobreak\smallskip
\noindent\begin{tabularx}{\linewidth}{@{}l L l@{}}
\toprule
Symbol & Description & Unit \\
\midrule
$\varepsilon$, $\varepsilon_m$ & heat / mass exchange effectiveness & --- \\
$\rho_{air}$, $\rho_{por}$ & air / porous-matrix density & kg\,m$^{-3}$ \\
$\theta_j$ & model parameter subject to sensitivity perturbation & --- \\
$\sigma_k$ & singular value of the sensitivity matrix & $^{\circ}$C \\
$\phi$ & matrix porosity & --- \\
$\omega$ & wheel rotation speed & rad\,s$^{-1}$ \\
\bottomrule
\end{tabularx}

\vspace{1em}
\noindent\textbf{Subscripts}\par\nobreak\smallskip
\noindent\begin{tabular}{@{}l l@{}}
\toprule
Sub & Meaning \\
\midrule
$ab$ & adsorption sector \\
$amb$ & ambient \\
$i, j$ & rotor zone index (proc / regen) \\
$in$, $out$ & inlet / outlet \\
$proc$, $regen$ & process / regeneration sector \\
$r$ & rotor matrix \\
\bottomrule
\end{tabular}

\vspace{1em}
\noindent\textbf{Abbreviations}\par\nobreak\smallskip
\noindent\begin{tabular}{@{}l l@{}}
\toprule
Abbr & Meaning \\
\midrule
AHU & air-handling unit \\
DH & dehumidified (wheel process-outlet) stream \\
HVAC & heating, ventilation and air conditioning \\
LDF & linear driving force \\
MA / OA / RA / SA / EA & mixed / outdoor / return / supply / exhaust air \\
PLC & programmable logic controller \\
TSL & triple-site Langmuir \\
YAML & the specification markup format \\
\bottomrule
\end{tabular}

%% Elsevier 생성형 AI 공시 — 정책상 References 직전 배치.
%% 문안 = 사용자 확정 (2026-08-06): "언어 다듬기로 한정" 안.
\section*{Declaration of generative AI and AI-assisted technologies in the
writing process}

During the preparation of this work the authors used a generative AI assistant
in order to improve the readability and language of the manuscript. After using
this tool, the authors reviewed and edited the content as needed and take full
responsibility for the content of the published article.

\bibliographystyle{elsarticle-num}
% local_additions = Zotero 미등록 임시 항목 (suto2026S220Manual) — 등록 후 정리
\bibliography{Zotero_YHJoo,local_additions}

\end{document}